\documentclass[11pt]{article}

\usepackage{amsmath,amssymb}

\newcommand{\ko}[1]{\left( #1 \right)}

\newcommand{\bmt}[1]{{{\mbox{\boldmath$ #1 $}}}}

\numberwithin{equation}{section}
\begin{document}

\quad 
\vspace{-3.5cm}

\begin{flushright}
\parbox{3cm}
{
{\bf October 2007}\hfill \\
UT-07-32 \hfill \\
 }
\end{flushright}

\vspace*{0.5cm}

\begin{center}
\Large\bf 
Refined Topological Vertex and Instanton Counting
\end{center}
\vspace*{0.7cm}
\centerline{\large 
Masato Taki
}
\begin{center}
\emph{Department of Physics, Faculty of Science, University of Tokyo,\\
Bunkyo-ku, Tokyo 113-0033, Japan.} \\
\vspace*{0.5cm}
{\tt tachyon@hep-th.phys.s.u-tokyo.ac.jp}
\end{center}

\vspace*{0.7cm}

\centerline{\bf Abstract} 

\vspace*{0.5cm}

It has been proposed recently that topological A-model string amplitudes for toric Calabi-Yau 3-folds in non self-dual graviphoton background can be caluculated by a diagrammatic method that is called the ``refined topological vertex''. We compute the extended A-model amplitudes for $SU(N)$-geometries using the proposed vertex. If the refined topological vertex is valid, these computations should give rise to the Nekrasov's partition functions of $\cal N$$=2$ $SU(N)$ gauge theories via the geometric engineering. In this article, we verify the proposal by confirming the equivalence between the refined A-model amplitude and the K-theoretic version of the Nekrasov's partition function by explicit computation.

\vspace*{1.0cm}

\vfill

\thispagestyle{empty}
\setcounter{page}{0}

\newpage

\section{Introduction}

The study of topological A-model strings on non-compact toric Calabi-Yau manifolds has been the important subject in the research of topological strings. Topological strings has provided insights into mathematics and nonperturbative dynamics of gauge and  string theory. 

On the one hand, it is in general very hard to caluculate the topological string partition functions exactly. However in some cases, various dualities enable us to simplify the caluculation of topological strings greatly and provide new perspectives \cite{review}. For example, the A-model partition function on the resolved conifold is given by the partition function of Chern-Simons theory on $S^3$ . This is the geometric transition between the resolved conifold and the deformed conifold \cite{Gopakumar:1998vy}\cite{Gopakumar:1998ii}\cite{Gopakumar:1998ki}. By generalizing this argument, an elegant technique for computing the A-model partition function on toric Calabi-Yau manifolds was formulated in \cite{Aganagic:2003db}. The formalism is called the topological vertex.

The mechanism of the geometric engineering is one way to study supersymmetric gauge theories using string theory and topological string \cite{engineering}. This approach tells us that we can caluculate the F-terms of various $\cal N$$=2$ $SU(N)$ gauge theories by using topological A-model strings on certain toric Calabi-Yau manifolds. The partition function of the A-model on the toric Calabi-Yau agrees with the Nekrasov's partition function of $\cal N$$=2$ $SU(N)$ gauge theory \cite{Nekrasov:2002qd}\cite{Iqbal:2003ix}\cite{Iqbal:2003zz}\cite{Eguchi:2003sj}\cite{Hollowood:2003cv}\cite{Zhou:2003}\cite{Eguchi:2003it}. Thus topological strings are useful tool to obtain insights into the nonperturbative dynamics of supersymmetric gauge theories. 

The Nekrasov's partition function in a constant self-dual graviphoton background contains
] one parameter which is corresponding to the value of the background field. The parameter is nothing but the topological string coupling constant in A-model side. On the other hand, we can perform the instanton caluclation in the more general background of non self-dual graviphoton configuration, and we get the K-theoretic version of the Nekrasov's partition function \cite{Nekrasov:2002qd}\cite{Flume:2002az}\cite{Bruzzo:2002xf}. Then the Nekrasov's partition function has one more parameter in addition to the self-dual graviphoton background. Hence it is natural to expect that there exists a 2-parameter extension of the topological vertex which will recover the K-theoretic answer. Few attempts were made for defining the 2-parameter extension of topological strings and formulating the algorithmical techniques to caluculate the extended partition function \cite{Hollowood:2003cv}\cite{Awata:2005fa}\cite{Zhou:2005}. Recently a refined topological vertex was proposed in \cite{Iqbal:2007ii}. In this article, we compute the refined topological A-model string partition function for the $SU(N)$ geomerties and check the equivalence of the refined partition function and the K-theoretic version of the Nekrasov's partition function.

This paper is organized as follows.
In section 2, we review the geometric engineering, the topological vertex and their 2-parameter extension. 
The refined A-model partition function for $SU(N)$ geomerties are caluculated in section 3.
Conclusions are found in section 4. In appendix A, we give brief introduction to Young diagrams, Schur functions, and the useful formulae for Schur functions.
In appendix B, a proof of a formula can be found.

%%%%%%%%%%%%%%%%%%%%%%%%%%%%%%%%%%%%%%%%%%%%%%%%%%%%%%%%%%%%%%%
\section{Topological Strings and Instanton Counting}
\label{sec:rev}
%%%%%%%%%%%%%%%%%%%%%%%%%%%%%%%%%%%%%%%%%%%%%%%%%%%%%%%%%%%%%%%%%

In this section, we will briefly review the idea of the geometric engineering, topological A-model strings, and the instanton counting.

%%%%%%%%%%%%%%%%%%%%%%%%%%%%%%%%%%%%%%%%%%%%%%%%%%%%%%%%%%%%%%%
\subsection{ Geometric Engineering and A-model}
%%%%%%%%%%%%%%%%%%%%%%%%%%%%%%%%%%%%%%%%%%%%%%%%%%%%%%%%%%%%%%%

Type IIA string theory compactified on a Calabi-Yau 3-fold yields an effective theory in transverse 4-dimensions. Especially, enhanced gauge symmetries arise from singular Calabi-Yau compactification. Thus in the field theory limit, appropriate Calabi-Yau compactifications provide effective gauge theories in 4-dimensions. This is the basic idea of the geometric engineering \cite{engineering}.
%${\mathbb P}^{1}$

Let us consider Type IIA compactified on a Calabi-Yau 3-fold $\mbox{\ M\ }$. The K\"{a}hler parameters of M are denoted by $
{\mathop t\nolimits_i } 
$. Then, the F-term of the effective theory is given by \cite{Antoniadis:1993ze}\cite{Bershadsky:1993cx}
%%%%%%%%%%%%%%%%%%
\begin{align}
\sum\limits_{g = 0}^\infty  {\int {\mathop d\nolimits^4 x\mathop d\nolimits^4 \theta \mathop W\nolimits^{2g} \mathop F\nolimits_g (\mathop t\nolimits_i )} } 
  = 
\int {\mathop d\nolimits^4 x\left[ {\mathop \tau \nolimits_{ij} \mathop F\nolimits_{\mu \nu }^i \mathop F\nolimits^{j\mu \nu }  + \sum\limits_{g = 1}^\infty  {\mathop F\nolimits_g (\mathop t\nolimits_i )\mathop {\mathop R\nolimits_ +  }\nolimits^2 \mathop {\mathop F\nolimits_ +  }\nolimits^{2g - 2} } } \right]} 
\end{align}
%%%%%%%%%%%%%%%%%%%%%%
Here, $W$ is $\mathop W\nolimits_{\mu \nu }  = \mathop F\nolimits_{\mu \nu }^ +   - \mathop R\nolimits_{\mu \nu \rho \sigma } \theta \mathop \sigma \nolimits^{\rho \sigma } \theta  +  \cdots $,
$
{\mathop F\nolimits_ +  }
$
is the self-dual part of the graviphoton field strength, 
$
{\mathop R\nolimits_ +  }
$
is the self-dual part of the Riemann tensor, and
$
{\mathop F\nolimits_{\mu \nu }^i }
$
is the $U(1)$ gauge field strength of the effective theory.
Notice that the 4-dimensional $U(1)$ gauge couplings are given by
%%%%%%%%%%%%%%%%
\begin{equation}
\mathop \tau \nolimits_{ij}  = \frac{{\mathop \partial \nolimits^2 }}{{\partial \mathop t\nolimits_i \partial \mathop t\nolimits_j }}\mathop F\nolimits_0 (\mathop t\nolimits_i )
\end{equation}
%%%%%%%%%%%%%%%%%
Hence the genus zero amplitudes of Type IIA strings $
\mathop F\nolimits_0 (\mathop t\nolimits_i )
$ give rise to the effective gauge couplings. This is the Seiberg-Witten theory \cite{Seiberg:1994rs} in Type IIA string theory set-up. The higher genus amplitudes $
\mathop F\nolimits_g (\mathop t\nolimits_i )
$ correspond to the graviphoton corrections to the gauge theory. They play an important role in the Nekrasov's partition function that gives a closed expression for the Seiberg-Witten prepotential \cite{Nekrasov:2002qd}\cite{Flume:2002az}\cite{Bruzzo:2002xf}.

Furthermore, the amplitudes of Type IIA strings $
\mathop F\nolimits_g (\mathop t\nolimits_i )
$ are identical with the topological A-model string amplitudes $
\mathop {\cal F}\nolimits_g (\mathop t\nolimits_i )
$ of ${\mbox{\ M\ }}$ which "count" the holomorphic maps from genus $g$ Riemann surfaces to a Calabi-Yau ${\mbox{\ M\ }}$ \cite{Antoniadis:1993ze}\cite{Bershadsky:1993cx}. The information of the partition function was encoded in the Gromov-Witten invariants. The generating function of these amplitudes is called the topological A-model string partition function
%%%%%%%%%%%%%%%%
\begin{equation}
Z =  \exp\ko{{\cal F}(\mathop g\nolimits_s ,\mathop t\nolimits_i )}  = \exp\ko{\sum\limits_{g = 0}^\infty  {g_s^{2g - 2} {\cal F}_g ( t_i )} } 
\end{equation}
%%%%%%%%%%%%%%%%%
Here $\mathop g\nolimits_s$ is the topological string couplins constant.

%%%%%%%%%%%%%%%%%%%%%%%%%%%%%%%%%%%%%%%%%%%%%%%%%%%%%%%%%%%%%%%
\subsection{ Gopakumar-Vafa Invariants}
%%%%%%%%%%%%%%%%%%%%%%%%%%%%%%%%%%%%%%%%%%%%%%%%%%%%%%%%%%%%%%%

The target space perspective tells us that we can reformulate A-model as BPS state counting problem. Let us consider M-theory lift of Type IIA on a Calabi-Yau, i.e. M-theory compactified on a Calabi-Yau times a circle. This set-up gives rise to an effective field theory in the transverse 5-dimension $
{\mathbb R}^{1,3}  \times \mathop {\mbox{\ S\ }}\nolimits^1 
$. The particles in the effective theory arise from M2 branes wrapping holomorphic curves of ${\mbox{\ M\ }}$. The mass and the charge $
\left( {\mathop j\nolimits_L ,\mathop j\nolimits_R } \right)
$ of the little group in 5-dimensions $
SO(4) = \mathop {SU(2)}\nolimits_L  \times \mathop {SU(2)}\nolimits_R 
$ characterise these BPS particles. The masses are given by $
\mathop m\nolimits_{\left( {\Sigma ,n} \right)}  = \mathop T\nolimits_\Sigma   + \frac{{2\pi in}}{{\mathop g\nolimits_s }}
$. Here $\mathop T\nolimits_\Sigma$ is the K\"{a}hler parameter of the curve class $\Sigma$ which M2 brane wraps, and $n$ is the momentum along $\mathop {\mbox{\ S\ }}\nolimits^1$. Therefore the mass (and cherge via BPS condition) is given by the curve class $\Sigma$ and  the momentum $n$. Integrating out these particles, we get the F-term of the effective theory \cite{Gopakumar:1998ii}%\cite{Gopakumar:1998jq}
\begin{eqnarray*}
{\cal F} &=& \sum\limits_{\Sigma  \in \mathop H\nolimits_2 (M,\mathbb Z)} {\sum\limits_{n \in Z} {\sum\limits_{\mathop j\nolimits_L ,\mathop j\nolimits_R } {\mathop N\nolimits_\Sigma ^{\left( {\mathop j\nolimits_L ,\mathop j\nolimits_R } \right)} \mathop {\log \det }\nolimits_{\left( {\mathop j\nolimits_L ,\mathop j\nolimits_R } \right)} \left( {\Delta  + \mathop {\mathop m\nolimits_{\left( {\Sigma ,n} \right)} }\nolimits^2  + 2{\mathop m\nolimits_{\left( {\Sigma ,n} \right)} }\mathop \sigma \nolimits_L \mathop F\nolimits_ +  } \right)} } } \\
 &=& \sum\limits_{\Sigma  \in \mathop H\nolimits_2 (M,\mathbb Z)} {\sum\limits_{k = 1}^\infty  {\sum\limits_{\mathop j\nolimits_L } {\mathop N\nolimits_\Sigma ^{\mathop j\nolimits_L } \mathop e\nolimits^{ - k\mathop T\nolimits_\Sigma  } \frac{{\mathop {Tr}\nolimits_{\mathop j\nolimits_L } \mathop {( - 1)}\nolimits^{\mathop \sigma \nolimits_L } \mathop e\nolimits^{ - 2k\mathop g\nolimits_s \mathop \sigma \nolimits_L } }}{{k\mathop {\left( {2\sinh (k\mathop g\nolimits_s /2)} \right)}\nolimits^2 }}} } } \\
 &=& \sum\limits_{\Sigma  \in \mathop H\nolimits_2 (M,\mathbb Z)} {\sum\limits_{k = 1}^\infty  {\sum\limits_{\mathop j\nolimits_L } {\mathop N\nolimits_\Sigma ^{\mathop j\nolimits_L } \mathop {( - 1)}\nolimits^{ - 2\mathop j\nolimits_L } \mathop e\nolimits^{ - k\mathop T\nolimits_\Sigma  } \frac{{\sum\limits_{l =  - \mathop j\nolimits_L }^{\mathop j\nolimits_L } {\mathop q\nolimits^{ - 2kl} } }}{{k\mathop {\left( {\mathop q\nolimits^{k/2}  - \mathop q\nolimits^{ - k/2} } \right)}\nolimits^2 }}} } } 
\end{eqnarray*}
Notice that the graviphoton expectation value gives topological string coupling $
\mathop F\nolimits_ +   = \mathop g\nolimits_s 
$ and we introduce $
q = \mathop e\nolimits^{ - \mathop g\nolimits_s } 
$. Changing representation basis of $\mathop {SU(2)}\nolimits_L $ so as to satisfy $\sum\limits_{\mathop j\nolimits_L } {\mathop N\nolimits_\Sigma ^{\mathop j\nolimits_L } (\sum\limits_{l =  - \mathop j\nolimits_L }^{\mathop j\nolimits_L } {\mathop q\nolimits^l } )}  = \sum\limits_{g = 0}^\infty  {\mathop n\nolimits_\Sigma ^g \mathop {( - 1)}\nolimits^g \mathop {(\mathop q\nolimits^{1/2}  - \mathop q\nolimits^{ - 1/2} )}\nolimits^{2g} } $, we get the following expression of the A-model partition function
\begin{equation}
{\cal F} = \log Z = \sum\limits_{\Sigma  \in \mathop H\nolimits_2 (M,\mathbb Z)} {\sum\limits_{k = 1}^\infty  {\sum\limits_{g = 0}^\infty  {\frac{{\mathop n\nolimits_\Sigma ^g }}{k}\mathop {(\mathop q\nolimits^{k/2}  - \mathop q\nolimits^{ - k/2} )}\nolimits^{2g - 2} \mathop e\nolimits^{ - \mathop T\nolimits_\Sigma  } } } } 
\end{equation}
Integer valued invariants ${\mathop n\nolimits_\Sigma ^g }$ which are defined as above are called "Gopakumar-Vafa invariants".

$
{\mathop N\nolimits_\Sigma ^{\left( {\mathop j\nolimits_L ,\mathop j\nolimits_R } \right)} }$ is the number of the wrapped M2-branes, and they are not invariant under the complex structure deformations of Calabi-Yau. Roughly speaking, this is the reason why the information encoded in the partition function is not the full degeneracies ${\mathop N\nolimits_\Sigma ^{\left( {\mathop j\nolimits_L ,\mathop j\nolimits_R } \right)} }$ but ${\mathop N\nolimits_\Sigma ^{\mathop j\nolimits_L } }$ which are summed over $\mathop {SU(2)}\nolimits_R $ charges as
\begin{equation}
\mathop N\nolimits_\Sigma ^{\mathop j\nolimits_L }  = \sum\limits_{\mathop j\nolimits_R } {\mathop {( - 1)}\nolimits^{ - 2\mathop j\nolimits_R } (2\mathop j\nolimits_R  + 1)\mathop N\nolimits_\Sigma ^{\left( {\mathop j\nolimits_L ,\mathop j\nolimits_R } \right)} } 
\end{equation}
However ${\mathop N\nolimits_\Sigma ^{\left( {\mathop j\nolimits_L ,\mathop j\nolimits_R } \right)} }$ themselves are invariants for non-compact Calabi-Yau since these Calabi-Yau 3-folds have no complex structure deformations \cite{Hollowood:2003cv}. Among them, local toric Calabi-Yau 3-folds are important ones. Hence we define an extended partition function that counts invariants ${\mathop N\nolimits_\Sigma ^{\left( {\mathop j\nolimits_L ,\mathop j\nolimits_R } \right)} }$ as follows
\begin{eqnarray*}
{\cal F} &=& \sum\limits_{\Sigma  \in \mathop H\nolimits_2 (M,\mathbb Z)} {\sum\limits_{n \in Z} {\sum\limits_{\mathop j\nolimits_L ,\mathop j\nolimits_R } {\mathop N\nolimits_\Sigma ^{\left( {\mathop j\nolimits_L ,\mathop j\nolimits_R } \right)} \mathop {\log \det }\nolimits_{\left( {\mathop j\nolimits_L ,\mathop j\nolimits_R } \right)} \left( {\Delta  + \mathop {\mathop m\nolimits_{\left( {\Sigma ,n} \right)} }\nolimits^2  + 2{\mathop m\nolimits_{\left( {\Sigma ,n} \right)} }\mathop \sigma \nolimits_L \left( {\mathop F\nolimits_ +   + \mathop F\nolimits_ -  } \right)} \right)} } } \\
 &=& \sum\limits_{\Sigma  \in \mathop H\nolimits_2 (M,Z)} {\sum\limits_{k = 1}^\infty  {\sum\limits_{\mathop j\nolimits_L ,\mathop j\nolimits_R } {\mathop N\nolimits_\Sigma ^{\left( {\mathop j\nolimits_L ,\mathop j\nolimits_R } \right)} \mathop {( - 1)}\nolimits^{ - 2\left( {\mathop j\nolimits_L  + \mathop j\nolimits_R } \right)} \mathop e\nolimits^{ - k\mathop T\nolimits_\Sigma  } \frac{{\left( {\sum\limits_{l =  - \mathop j\nolimits_L }^{\mathop j\nolimits_L } {\mathop {\left( {tq} \right)}\nolimits^{ - 2kl} } } \right)\left( {\sum\limits_{m =  - \mathop j\nolimits_R }^{\mathop j\nolimits_R } {\mathop {\left( {\frac{t}{q}} \right)}\nolimits^{ - 2km} } } \right)}}{{k\left( {\mathop t\nolimits^{k/2}  - \mathop t\nolimits^{ - k/2} } \right)\left( {\mathop q\nolimits^{k/2}  - \mathop q\nolimits^{ - k/2} } \right)}}} } } 
\end{eqnarray*}
Here $q = \mathop e\nolimits^{\mathop F\nolimits_ +  } $ and $t = \mathop e\nolimits^{\mathop F\nolimits_ -  }$.

The question now arises; how to compute these partition functions for non-compact Calabi-Yau. In the case of toric Calabi-Yau 3-folds, the answer can be found in a diagrammatic methods named the topological vertex. Before we turn to the discussion of topological vertex, it will be useful to take a look at the instanton counting of $\cal N$$=2$ gauge theory. Hence in the next section, we discuss the Nekrasov's partition function of $\cal N$$=2$ gauge theory. We will come back to the discussion of the topological vertex later.

%%%%%%%%%%%%%%%%%%%%%%%%%%%%%%%%%%%%%%%%%%%%%%%%%%%%%%%%%%%%%%%
\subsection{Instanton Counting of \bmt{{\cal N}=2} Gauge Theories}
%%%%%%%%%%%%%%%%%%%%%%%%%%%%%%%%%%%%%%%%%%%%%%%%%%%%%%%%%%%%%%%

Instanton calculation of $\cal N$$=2$ gauge theories in 4- and 5-dimensions has been developed by Nekrasov \cite{Nekrasov:2002qd}. He found that the instanton coefficients of the Seiberg-Witten prepotential are summed up to a closed form, and he provided the combinatorical expression of this generating function. We call it the Nekrasov's partition function. His conjectual observation was mathematically verified by Nekrasov-Okounkov \cite{Nekrasov:2003rj}, Nakajima-Yoshioka \cite{Nakajima:2003}, and Braverman \cite{Braverman}.

Take an $\cal N$$=2$ $SU(N)$ supersymmetric pure Yang-Mills theory for example. Muiti-instanton calculation involves an integral over the ADHM moduli space. It is in general very hard to carry out the caluculation. However we can formulate the muiti-instanton calculation of $\cal N$$=2$ $SU(N)$ supersymmetric gauge theory as integrals of equivariant closed forms. Let us consider the following partition function of $\cal N$$=2$ $SU(N)$ supersymmetric pure Yang-Mills theory
\begin{align}
\mathop Z\nolimits^{inst.} (\vec a,\Lambda ) = \sum\limits_{k = 1}^\infty  {\mathop \Lambda \nolimits^{2Nk} \mathop Z\nolimits^k (\vec a)} \nonumber
\end{align}
Here $\vec a$ ia the Coulomb moduli, $\Lambda$ is the dynamical scale, and $\mathop Z\nolimits^k (\vec a)$ is a k-instanton contribution. By deforming the theory by torus action on the moduli space, we can give the partition function as an integral of equivariant differential
\begin{align}
\mathop Z\nolimits^{k}  = \int\limits_{{\cal M}(N,k)} {{\cal D}\mu \mathop e\nolimits^{ - Q\Psi } } 
\end{align}
where ${\cal M}(N,k)$ is the ADHM moduli space of k-instantons and $Q$ is the BRST operator. It is known that the BRST operator is an equivariant differential for torus action $T =\mathop {U(k)} \times \mathop {U(1)}\nolimits^{N - 1}  \times \mathop {U(1)}\nolimits^2 $ on the moduli space. 
Here $\mathop {U(1)}\nolimits^{2}$ is the rotation groups of complex plane $\mathop {\mathbb R}\nolimits^4  = \mathop {\mathbb C}\nolimits^2 $ and their weights provide deformation parameters $\mathop \epsilon \nolimits_i$. Then we can apply the localization formula 
\begin{equation}
\mathop Z\nolimits^{k }  = \sum\limits_{\mathop p\nolimits_0 } {\frac{1}{{\sqrt {\det \mathop {\cal L}\nolimits_{\mathop p\nolimits_0 } } }}} 
\end{equation}
Here $\mathop p\nolimits_0$ are isolated fixed points of the torus action and $\mathop {\cal L}\nolimits_{\mathop p\nolimits_0 }$ is the Lie derivative acting on the tangent moduli space $T{\cal M}(N,k)$
. It is known that the fixed points of $T$-action are uniquely specified by $N$ Young diagrams $(\mathop \mu \nolimits_1 , \cdots ,\mathop \mu \nolimits_N )$.
Then we have to know the weights $\det \mathop {\cal L}\nolimits_{\mathop p\nolimits_0 }$ of $T$-action on the tangent moduli space $T{\cal M}(N,k)$ for the purpose of multi-instanton caluculus. The weights were caluculated in \cite{nakajima}\cite{Nekrasov:2002qd}\cite{Flume:2002az}\cite{Bruzzo:2002xf} and the explicit expression is given by
\begin{align}
\label{nek4d}
\mathop Z\nolimits^{inst.} (\mathop \epsilon \nolimits_1 ,\mathop \epsilon \nolimits_2 ,\vec a,\Lambda )
 &= \sum\limits_{\vec \mu } {\mathop \Lambda \nolimits^{2N\left| {\vec \mu } \right|} \prod\limits_{a,b = 1}^N {\prod\limits_{s \in \mathop \mu \nolimits_a } {\frac{1}{{\mathop a\nolimits_b  - \mathop a\nolimits_a  - \mathop \epsilon \nolimits_1 \mathop l\nolimits_{\mathop \mu \nolimits_b } (s) + \mathop \epsilon \nolimits_2 (\mathop a\nolimits_{\mathop \mu \nolimits_a } (s) + 1)}}} } } \nonumber\\
&\times \prod\limits_{t \in \mathop \mu \nolimits_b } {\frac{1}{{\mathop a\nolimits_b  - \mathop a\nolimits_a  + \mathop \epsilon \nolimits_1 (\mathop l\nolimits_{\mathop \mu \nolimits_a } (s) + 1) - \mathop \epsilon \nolimits_2 \mathop a\nolimits_{\mathop \mu \nolimits_b } (s)}}} 
\end{align}
Nekrasov claimed that the partition function (\ref{nek4d}) leads to the Seiberg-Witten prepotential after eliminating the deformation parameter $\epsilon$ as follows
\begin{align}
{\mathop \epsilon \nolimits_1}{\mathop \epsilon \nolimits_2} \log \mathop Z\nolimits^{inst.} (\mathop \epsilon \nolimits_1 ,\mathop \epsilon \nolimits_2 ,\vec a,\Lambda ) = \mathop {\cal F}\nolimits_{SW}^{inst.} (\vec a,\Lambda ) + O(\mathop \epsilon \nolimits_1,\mathop \epsilon \nolimits_2)
\end{align}
This conjecture was proved by using the thermodynamical limit of the random partition \cite{Nekrasov:2003rj}, the blow-up equation \cite{Nakajima:2003}, and \cite{Braverman}.

We can lift it to the 5-dimensional gauge theory result
\begin{align}
\label{nek5}
\mathop Z\nolimits_{5D}^{inst.} (t ,q ,\vec a,\Lambda ,\beta ) = \sum\limits_{\vec \mu } {\mathop {(\beta \Lambda )}\nolimits^{2N\left| {\vec \mu } \right|} \prod\limits_{a,b = 1}^N {\prod\limits_{s \in \mathop \mu \nolimits_a } {\frac{1}{{1 - \mathop Q\nolimits_{ba} \mathop t\nolimits^{\mathop l\nolimits_{\mathop \mu \nolimits_b } (s)} \mathop q\nolimits^{\mathop a\nolimits_{\mathop \mu \nolimits_a } (s) + 1} }}\prod\limits_{t \in \mathop \mu \nolimits_b } {\frac{1}{{1 - \mathop Q\nolimits_{ba} \mathop t\nolimits^{ - \mathop l\nolimits_{\mathop \mu \nolimits_a } (s) - 1} \mathop q\nolimits^{ - \mathop a\nolimits_{\mathop \mu \nolimits_b } (s)} }}} } } } 
\end{align}
Here $\beta $ is the radius of the compact fifth dimension $S^1$.
Let us choose the deformation parameter $\epsilon$ as $\hbar  = \mathop \epsilon \nolimits_1  =  - \mathop \epsilon \nolimits_2 $. In \cite{Eguchi:2003sj}\cite{Iqbal:2003zz}\cite{Hollowood:2003cv} the partition function was reproduced from the string calculation via the geometric engineering
\begin{align}
\label{nek5dA}
\mathop Z\nolimits_{5D}^{Nek. SU(N)} (\hbar , - \hbar ,\vec a,\Lambda ,\beta ) = \mathop Z\nolimits^{A - model, SU(N)} (\hbar  = \mathop g\nolimits_s ,\mathop t\nolimits_i )
\end{align}
and they verified the interpretation in \cite{Nekrasov:2002qd} that $\hbar $ expansion is nothing but the genus expansion of the string partition function. Notice that the Coulomb moduli $\vec a$ and the dynamical scale $\Lambda $ are engineered from the K\"{a}hler parameters of the Calabi-Yau. We review the results (\ref{nek5dA}) for $SU(2)$ theory later.

Thus it is natural to expect that there is a refinement of string theory to engineer Nekrasov' partition function for the general case $\mathop \epsilon \nolimits_1  \ne   - \mathop \epsilon \nolimits_2 $.
In this paper we calculate the K-theoretic partition function (\ref{nek5}) via the refined topological vertex and show that the refined A-model of \cite{Iqbal:2007ii} reproduces the correct results.

%%%%%%%%%%%%%%%%%%%%%%%%%%%%%%%%%%%%%%%%%%%%%%%%%%%%%%%%%%%%%%%
\subsection{ Topological Vertex and its refinement}
%%%%%%%%%%%%%%%%%%%%%%%%%%%%%%%%%%%%%%%%%%%%%%%%%%%%%%%%%%%%%%%

It is known that we can compute the topological A-model string amplitudes for toric Calabi-Yau 3-folds by using the topological vertex \cite{Aganagic:2003db}. The topological vertex is the Feymnan-rules like technique which arises from the geometric transition between A-model and Chern-Simons gauge theory. The Feymnan diagrams, the vertices of diagrams, the momentun, and the propagators are corresponding to the toric web-diagrams, the tri-valent vertices $
\mathop C\nolimits_{\mathop \mu \nolimits_1 \mathop \mu \nolimits_2 \mathop \mu \nolimits_3 } 
$, Young diagrams $\mu $, and the weights $
\mathop {( - 1)}\nolimits^{(n + 1)\left| \mu  \right|} 
\mathop { \mathop e\nolimits^{ - T} }\nolimits^{\left| \mu  \right|} \mathop q\nolimits^{ - \frac{{n\mathop \kappa \nolimits_\mu  }}{2}} 
$, respectively. Here, $T$ is the K\"{a}hler parameter for the 2-cycle corresponding to the line of the web-diagram, $\mu$ is the Young diagram which propagates along the line. The framing number $n$ is determined by the toric diagram. The vertex is expressed using the Schur functions
\begin{equation}
\mathop C\nolimits_{\lambda \mu \nu } (q) = \mathop q\nolimits^{\mathop \kappa \nolimits_\mu  /2} \mathop s\nolimits_{\mathop \nu \nolimits^t } (\mathop q\nolimits^{ - \rho } )\sum\limits_\eta  {\mathop s\nolimits_{\mathop \lambda \nolimits^t /\eta } (\mathop q\nolimits^{ - \nu  - \rho } )\mathop s\nolimits_{\mu /\eta } (\mathop q\nolimits^{ - \mathop \nu \nolimits^t  - \rho } )} 
\end{equation}
%%%%%%%%%%%%%%%%%%%%%%%%%%%%%%%%%%%%%%%%%%%%%%%%%%%%%%%%%%%%
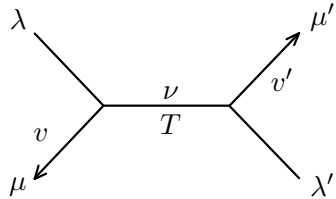
\begin{figure}[h]
\begin{center}
%WinTpicVersion3.08
\unitlength 0.1in
\begin{picture}( 22.9300,  9.5500)( -2.3100, -9.8700)
% LINE 1 0 3 0
% 4 619 226 980 603 980 603 619 984
% 
\special{pn 13}%
\special{pa 620 226}%
\special{pa 980 604}%
\special{fp}%
\special{pa 980 604}%
\special{pa 620 984}%
\special{fp}%
% LINE 1 0 3 0
% 6 977 603 1640 603 1640 603 2001 226 1637 603 2004 984
% 
\special{pn 13}%
\special{pa 978 604}%
\special{pa 1640 604}%
\special{fp}%
\special{pa 1640 604}%
\special{pa 2002 226}%
\special{fp}%
\special{pa 1638 604}%
\special{pa 2004 984}%
\special{fp}%
% STR 2 0 3 0
% 3 579 149 579 202 3 0
% $\lambda$
\put(5.7900,-2.0200){\makebox(0,0)[rb]{$\lambda$}}%
% STR 2 0 3 0
% 3 579 1023 579 1076 3 0
% $\mu$
\put(5.7900,-10.7600){\makebox(0,0)[rb]{$\mu$}}%
% STR 2 0 3 0
% 3 2062 160 2062 212 2 0
% $\mu' $
\put(20.6200,-2.1200){\makebox(0,0)[lb]{$\mu' $}}%
% STR 2 0 3 0
% 3 2062 1023 2062 1076 2 0
% $\lambda'$
\put(20.6200,-10.7600){\makebox(0,0)[lb]{$\lambda'$}}%
% STR 2 0 3 0
% 3 1332 484 1332 536 5 0
% $\nu$
\put(13.3200,-5.3600){\makebox(0,0){$\nu$}}%
% STR 2 0 3 0
% 3 1332 646 1332 699 5 0
% $T$
\put(13.3200,-6.9900){\makebox(0,0){$T$}}%
% LINE 1 0 3 0
% 4 640 917 619 987 619 987 680 959
% 
\special{pn 13}%
\special{pa 640 918}%
\special{pa 620 988}%
\special{fp}%
\special{pa 620 988}%
\special{pa 680 960}%
\special{fp}%
% LINE 1 0 3 0
% 4 2004 223 1938 255 2004 223 1977 285
% 
\special{pn 13}%
\special{pa 2004 224}%
\special{pa 1938 256}%
\special{fp}%
\special{pa 2004 224}%
\special{pa 1978 286}%
\special{fp}%
% STR 2 0 3 0
% 3 654 704 654 756 5 0
% $v$
\put(6.5400,-7.5600){\makebox(0,0){$v$}}%
% STR 2 0 3 0
% 3 1857 474 1857 526 2 0
% $v '$
\put(18.5700,-5.2600){\makebox(0,0)[lb]{$v '$}}%
\end{picture}%
\end{center}
\caption{The toric diagram obtained by gluing the vertices $C_{\lambda \mu \nu}$ and $C_{\lambda '\mu '\nu^t }$}
\label{glue}
\end{figure}
%%%%%%%%%%%%%%%%%%%%%%%%%%%%%%%%%%%%%%%%%%%%%%%%%%%%%%%%%%%%%
The vertices in Fig.\ref{glue} are glued as
\begin{equation}
\sum\limits_\nu  {\mathop C\nolimits_{\lambda \mu \nu } (q)\mathop {( - 1)}\nolimits^{(n + 1)\left| \nu  \right|} \mathop q\nolimits^{\mathop { - n\kappa }\nolimits_\nu  /2} \mathop e\nolimits^{ - T\left| \nu  \right|} \mathop C\nolimits_{\lambda '\mu '\mathop \nu \nolimits^t } (q)} 
\end{equation}
where the framing number $n$ is given by $n = v' \wedge v = \mathop {v'}\nolimits_1 \mathop v\nolimits_2  - \mathop v\nolimits_1 \mathop {v'}\nolimits_2 $.

%%%%%%%%%%%%%%%%%%%%%%%%%%%%%%%%%%%%%%%%%%%%%%%%%
\begin{figure}[htbp]
\begin{center}
%WinTpicVersion3.08
\unitlength 0.1in
\begin{picture}( 19.1200, 14.2200)( -0.0400,-16.9600)
% LINE 1 0 3 0
% 12 656 452 960 751 960 751 1571 751 1571 751 1874 444 960 755 960 1374 960 1374 1578 1374 1578 1374 1578 747
% 
\special{pn 13}%
\special{pa 656 452}%
\special{pa 960 752}%
\special{fp}%
\special{pa 960 752}%
\special{pa 1572 752}%
\special{fp}%
\special{pa 1572 752}%
\special{pa 1874 444}%
\special{fp}%
\special{pa 960 756}%
\special{pa 960 1374}%
\special{fp}%
\special{pa 960 1374}%
\special{pa 1578 1374}%
\special{fp}%
\special{pa 1578 1374}%
\special{pa 1578 748}%
\special{fp}%
% LINE 1 0 3 0
% 6 1578 1377 1878 1681 964 1370 964 1370 964 1370 649 1688
% 
\special{pn 13}%
\special{pa 1578 1378}%
\special{pa 1878 1682}%
\special{fp}%
\special{pa 964 1370}%
\special{pa 964 1370}%
\special{fp}%
\special{pa 964 1370}%
\special{pa 650 1688}%
\special{fp}%
% STR 2 0 3 0
% 3 634 414 634 452 3 0
% $\phi $
\put(6.3400,-4.5200){\makebox(0,0)[rb]{$\phi $}}%
% STR 2 0 3 0
% 3 1901 406 1901 444 2 0
% $\phi $
\put(19.0100,-4.4400){\makebox(0,0)[lb]{$\phi $}}%
% STR 2 0 3 0
% 3 1908 1658 1908 1696 1 0
% $\phi $
\put(19.0800,-16.9600){\makebox(0,0)[lt]{$\phi $}}%
% STR 2 0 3 0
% 3 626 1654 626 1692 4 0
% $\phi $
\put(6.2600,-16.9200){\makebox(0,0)[rt]{$\phi $}}%
% STR 2 0 3 0
% 3 903 1082 903 1121 3 0
% $\mu_1$
\put(9.0300,-11.2100){\makebox(0,0)[rb]{$\mu_1$}}%
% STR 2 0 3 0
% 3 1229 1396 1229 1434 1 0
% $\mu_2 $
\put(12.2900,-14.3400){\makebox(0,0)[lt]{$\mu_2 $}}%
% STR 2 0 3 0
% 3 1639 1076 1639 1114 2 0
% $\mu_3 $
\put(16.3900,-11.1400){\makebox(0,0)[lb]{$\mu_3 $}}%
% STR 2 0 3 0
% 3 1223 654 1223 692 2 0
% $\mu_4 $
\put(12.2300,-6.9200){\makebox(0,0)[lb]{$\mu_4 $}}%
% STR 2 0 3 0
% 3 1270 811 1270 850 5 0
% $Q_B $
\put(12.7000,-8.5000){\makebox(0,0){$Q_B $}}%
% STR 2 0 3 0
% 3 1076 1031 1076 1070 5 0
% $Q_F $
\put(10.7600,-10.7000){\makebox(0,0){$Q_F $}}%
% STR 2 0 3 0
% 3 1470 1031 1470 1070 5 0
% $Q_F $
\put(14.7000,-10.7000){\makebox(0,0){$Q_F $}}%
% STR 2 0 3 0
% 3 1280 1232 1280 1270 5 0
% $Q_B $
\put(12.8000,-12.7000){\makebox(0,0){$Q_B $}}%
\end{picture}%
\end{center}
\caption{The local Hirzebruch surface which is a line bundle over 
$\mathop {\mathbb P}^{1}  \times {\mathbb P}^{1} $}
\label{hirz}
\end{figure}
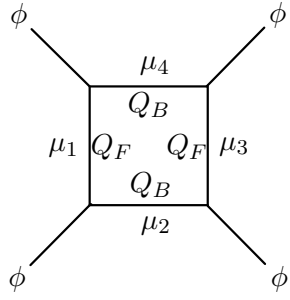
The local Hirzebruch surface $\mathop {\mathbb F}\nolimits_0 $ 
$\mathop = C({{\mathbb P}^{1}  \times {\mathbb P}^{1}} )$ 
is a good example to illustlate the topological vertex calculation. This toric Calabi-Yau 3-fold is the typical $SU(2)$ geometry that engineers $SU(2)$ pure super Yang-Mills theory. The toric diagram is given by Fig.\ref{hirz}, and we can easily check that the framing numbers associated to the four internal lines are all 1. Appling the topological vertex to Fig.\ref{hirz}, we get the following partition function
\begin{eqnarray*}
\mathop Z\nolimits^{\mathop F\nolimits_0 } (\mathop Q\nolimits_F ,\mathop Q\nolimits_B ) &=& \sum\limits_{\mathop \mu \nolimits_1 ,\mathop \mu \nolimits_2 ,\mathop \mu \nolimits_3 ,\mathop \mu \nolimits_4 } {\mathop {\mathop Q\nolimits_F }\nolimits^{\left| {\mathop \mu \nolimits_1 } \right| + \left| {\mathop \mu \nolimits_3 } \right|} \mathop {\mathop Q\nolimits_B }\nolimits^{\left| {\mathop \mu \nolimits_2 } \right| + \left| {\mathop \mu \nolimits_4 } \right|} \mathop q\nolimits^{\mathop { - \kappa }\nolimits_{\mathop \mu \nolimits_1 } /2\mathop { + \kappa }\nolimits_{\mathop \mu \nolimits_2 } /2\mathop { - \kappa }\nolimits_{\mathop \mu \nolimits_3 } /2\mathop { - \kappa }\nolimits_{\mathop \mu \nolimits_4 } /2} } \\
&& \times \mathop C\nolimits_{\phi \mathop \mu \nolimits_1 \mathop {\mathop \mu \nolimits_4 }\nolimits^t } \mathop C\nolimits_{\phi \mathop {\mathop \mu \nolimits_2 }\nolimits^t \mathop {\mathop \mu \nolimits_1 }\nolimits^t } \mathop C\nolimits_{\mathop \mu \nolimits_2 \phi \mathop \mu \nolimits_3 } \mathop C\nolimits_{\phi \mathop \mu \nolimits_4 \mathop {\mathop \mu \nolimits_3 }\nolimits^t } \\&=& \sum\limits_{\mathop \mu \nolimits_2 ,\mathop \mu \nolimits_4 } {\mathop {\mathop Q\nolimits_B }\nolimits^{\left| {\mathop \mu \nolimits_2 } \right| + \left| {\mathop \mu \nolimits_4 } \right|} \mathop q\nolimits^{\mathop { + \kappa }\nolimits_{\mathop \mu \nolimits_2 } /2\mathop { - \kappa }\nolimits_{\mathop \mu \nolimits_4 } /2} \mathop K\nolimits_{\mathop \mu \nolimits_4 \mathop \mu \nolimits_2 } (\mathop Q\nolimits_F )\mathop K\nolimits_{\mathop {\mathop \mu \nolimits_2 }\nolimits^t \mathop {\mathop \mu \nolimits_4 }\nolimits^t } (\mathop Q\nolimits_F )} 
\end{eqnarray*}
$K_{\mu \nu }$ is defined as follows
\begin{eqnarray*}
\mathop K\nolimits_{\mu \nu }  &=& \sum\limits_\lambda  {\mathop {\mathop Q\nolimits_F }\nolimits^{\left| \lambda  \right|} \mathop q\nolimits^{ - \mathop \kappa \nolimits_\lambda  /2} \mathop C\nolimits_{\phi \lambda \mathop \mu \nolimits^t } \mathop C\nolimits_{\mathop \nu \nolimits^t \mathop \lambda \nolimits^t \phi } } \\
 &=& \mathop s\nolimits_{\mathop \mu \nolimits^t } (\mathop q\nolimits^{ - \rho } )\mathop s\nolimits_\nu  (\mathop q\nolimits^{ - \rho } )\sum\limits_\lambda  {\mathop {\mathop Q\nolimits_F }\nolimits^{\left| \lambda  \right|} \mathop s\nolimits_\lambda  (\mathop q\nolimits^{ - \mu  - \rho } )\mathop s\nolimits_\lambda  (\mathop q\nolimits^{ - \mathop \nu \nolimits^t  - \rho } )} 
\\
 &=& \mathop q\nolimits^{\mathop {\left\| \mu  \right\|}\nolimits^2 /2 + || { \nu^t } ||^2 /2} \mathop {\tilde Z}\nolimits_{\mathop \mu \nolimits^t } (q)\mathop {\tilde Z}\nolimits_\nu  (q)\prod\limits_{i,j = 1}^\infty  {\frac{1}{{1 - \mathop Q\nolimits_F \mathop q\nolimits^{ - \mathop \mu \nolimits_i  - \mathop {\mathop \nu \nolimits^t }\nolimits_j  + i + j - 1} }}} 
\end{eqnarray*}
where we use the relation $\mathop s\nolimits_\mu  (\mathop q\nolimits^{ - \rho } ) = \mathop q\nolimits^{|| {\mu^t } ||^2 /2} \prod\limits_{s \in \mu } {\mathop {(1 - \mathop q\nolimits^{\mathop h\nolimits_\mu  (s)} )}\nolimits^{ - 1} }  = \mathop q\nolimits^{|| { \mu^t } ||^2 /2} \mathop {\tilde Z}\nolimits_\mu  (q)$
and formula (\ref{sum1}). Let us separate out the perturbative contributions as
\begin{equation}
\mathop Z\nolimits^{\mathop {\mathbb F}\nolimits_0 } \left( {\mathop Q\nolimits_B ,\mathop Q\nolimits_F } \right) = \mathop Z\nolimits_{pert.}^{\mathop {\mathbb F}\nolimits_0 } \left( {\mathop Q\nolimits_F } \right)\mathop Z\nolimits_{inst.}^{\mathop {\mathbb F}\nolimits_0 } \left( {\mathop Q\nolimits_B ,\mathop Q\nolimits_F } \right)
\end{equation}
\begin{equation}
\mathop Z\nolimits_{pert.}^{\mathop {\mathbb F}\nolimits_0 } \left( {\mathop Q\nolimits_F } \right) \equiv \mathop {\mathop K\nolimits_{\phi \phi } \left( {\mathop Q\nolimits_F } \right)}\nolimits^2  = \mathop {\left[ {\prod\limits_{i,j = 1}^\infty  {\frac{1}{{1 - \mathop Q\nolimits_F \mathop q\nolimits^{i + j - 1} }}} } \right]}\nolimits^2 
\end{equation}
Then, we get the A-model partition function corresponding to the nonperturbative part of the Nekrasov's partition function
\begin{equation}
\mathop Z\nolimits_{inst.}^{\mathop {\mathbb F}\nolimits_0 }  = \sum\limits_{\mu ,\nu } {\mathop {\mathop Q\nolimits_B }\nolimits^{\left| \mu  \right| +| \nu|} \mathop q\nolimits^{\mathop {\left\| \mu  \right\|}\nolimits^2  + || {\nu^t } ||^2 } \mathop {\tilde Z}\nolimits_\mu  (q)\mathop {\tilde Z}\nolimits_{\mathop \mu \nolimits^t } (q)\mathop {\tilde Z}\nolimits_\nu  (q)\mathop {\tilde Z}\nolimits_{\mathop \nu \nolimits^t } (q)\mathop {\left[ {\prod\limits_{i,j = 1}^\infty  {\frac{{1 - \mathop Q\nolimits_F \mathop q\nolimits^{ + i + j - 1} }}{{1 - \mathop Q\nolimits_F \mathop q\nolimits^{ - \mathop \mu \nolimits_i  - \mathop {\mathop \nu \nolimits^t }\nolimits_j  + i + j - 1} }}} } \right]}\nolimits^2 } 
\end{equation}
In fact, appling the formula (\ref{equal}) for the special case we can show that the above result is identical with the Nekrasov's partition function of the $SU(2)$ Yang-Mills theory (\ref{nek5}) for $t=q$. The identifications of parameters are given by
\begin{equation}
\mathop Q\nolimits_B  = (\beta \Lambda )^4  , \quad \mathop Q\nolimits_F  = \mathop e\nolimits^{2\beta a} 
\end{equation}

%%%%%%%%%%%%%%%%%%%%%%%%%%%%%%%%%%%%%%%%%%%%%%%%%%%%%%%%
\begin{figure}[htbp]
\begin{center}
%WinTpicVersion3.08
\unitlength 0.1in
\begin{picture}( 17.3300, 10.2100)(  0.3900,-10.5800)
% LINE 1 0 3 0
% 6 1302 216 1302 746 1302 746 1722 1021 1298 746 878 1021
% 
\special{pn 13}%
\special{pa 1302 216}%
\special{pa 1302 746}%
\special{fp}%
\special{pa 1302 746}%
\special{pa 1722 1022}%
\special{fp}%
\special{pa 1298 746}%
\special{pa 878 1022}%
\special{fp}%
% LINE 1 0 3 0
% 4 1267 365 1337 365 1267 392 1337 392
% 
\special{pn 13}%
\special{pa 1268 366}%
\special{pa 1338 366}%
\special{fp}%
\special{pa 1268 392}%
\special{pa 1338 392}%
\special{fp}%
% STR 2 0 3 0
% 3 930 848 930 887 3 0
% $q$
\put(9.3000,-8.8700){\makebox(0,0)[rb]{$q$}}%
% STR 2 0 3 0
% 3 1740 848 1740 887 3 0
% $t$
\put(17.4000,-8.8700){\makebox(0,0)[rb]{$t$}}%
% STR 2 0 3 0
% 3 849 1013 849 1051 4 0
% $\lambda$
\put(8.4900,-10.5100){\makebox(0,0)[rt]{$\lambda$}}%
% STR 2 0 3 0
% 3 1772 1021 1772 1058 1 0
% $\mu$
\put(17.7200,-10.5800){\makebox(0,0)[lt]{$\mu$}}%
% STR 2 0 3 0
% 3 1299 85 1299 122 5 0
% $\nu$
\put(12.9900,-1.2200){\makebox(0,0){$\nu$}}%
\end{picture}%
\end{center}
\caption{The refined topological vertex $\mathop C\nolimits_{\lambda \mu \nu } (t,q)$}
\label{refined}
\end{figure}
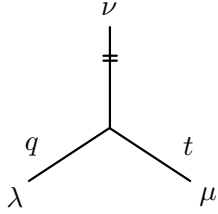
Recently, the topological vertex formalism for the refined partition functions has been proposed in \cite{Iqbal:2007ii} via melting crystal picture of the topological vertex. We call it the refined topological vertex. It was claimed that the refined topological vertex is constructed so as to engineer the K-theoretic version of the Nekrasov's partition function. We verify this claim in the next section. The proposal of \cite{Iqbal:2007ii} is as follows: the refined vertex corresponding to Fig.\ref{refined} is given by
\begin{equation}
\mathop C\nolimits_{\lambda \mu \nu } (t,q) = \mathop {\left( {\frac{q}{t}} \right)}\nolimits^{\frac{{\left\| \mu  \right\|^2  + \left\| \nu  \right\|^2 }}{2}} \mathop t\nolimits^{\frac{{\mathop \kappa \nolimits_\mu  }}{2}} \mathop P\nolimits_{\mathop \nu \nolimits^t } (\mathop t\nolimits^{ - \rho } ;q,t)\sum\limits_\eta  {\mathop {\left( {\frac{q}{t}} \right)}\nolimits^{\frac{{\left| \eta  \right| + \left| \lambda  \right| - \left| \mu  \right|}}{2}} \mathop s\nolimits_{\mathop \lambda \nolimits^t /\eta } (\mathop t\nolimits^{ - \rho } \mathop q\nolimits^{ - \nu } )\mathop s\nolimits_{\mu /\eta } (\mathop t\nolimits^{ - \mathop \nu \nolimits^t } \mathop q\nolimits^{ - \rho } )} 
\end{equation}
and we glue the "$t$-edge" and the "$q$-edge" with weight
\begin{equation}
\mathop f\nolimits_\mu  (t,q) = \mathop {( - 1)}\nolimits^{\left| \mu  \right|} \mathop t\nolimits^{n(\mu )} \mathop q\nolimits^{ - n(\mu ^t )} 
\end{equation}

The purpose of this article is to confirm that the refined vertex for $SU(N)$ geometry engineers the K-theoretic version of the Nekrasov's partition function. The refined partition functions for the $SU(2)$ and $SU(3)$ geometries and their blow-up were computed in \cite{Iqbal:2007ii}. Hence in the next section, we generalize their discussion to the general $SU(N)$ geometries and their blow-up. As the result, we propose that we shoud modifiy the framing factors in order to engineer the Nekrasov's results.

%%%%%%%%%%%%%%%%%%%%%%%%%%%%%%%%%%%%%%%%%%%%%%%%%%%%%%%%%%%%%%%%%%
\section{Refined A-model Amplitudes and Nekrasov's partition functions}
\label{ssec:3}
%%%%%%%%%%%%%%%%%%%%%%%%%%%%%%%%%%%%%%%%%%%%%%%%%%%%%%%%%%%%%%%%%%

In this section, we compute the refined partition function for $SU(N)$ geometry
 via refined topological vertex. The important point of the result in this section is that these refined partition functions are the same as the K-theoretic version of the Nekrasov's partition functions under the little modification of the framing factor. This result verifies the proposal of the refined topological vertex.

%%%%%%%%%%%%%%%%%%%%%%%%%%%%%%%%%%%%%%%%%%%%%%%%%%%%%%%%%%%%%%%%%%
%\subsection{$\cal N$$=2$ $SU(N)$ Super Yang-Mills}
%\label{ssec:3.1}
\subsection{\bmt{{\cal N}=2} \bmt{SU(N)} Super Yang-Mills}
\label{ssec:3.1}
%%%%%%%%%%%%%%%%%%%%%%%%%%%%%%%%%%%%%%%%%%%%%%%%%%%%%%%%%%%%%%%%%%
\subsubsection{A-model Partition Function}

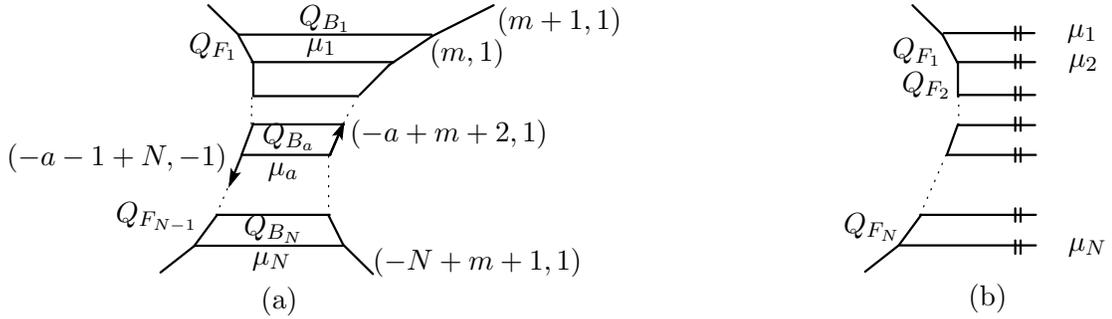
\begin{figure}[htbp]
\begin{center}
%WinTpicVersion3.08
\unitlength 0.1in
\begin{picture}( 55.5700, 14.8000)( -3.3700,-16.3500)
% LINE 1 0 3 0
% 2 5112 597 5112 667
% 
\special{pn 13}%
\special{pa 5112 598}%
\special{pa 5112 668}%
\special{fp}%
% LINE 1 0 3 0
% 2 5140 597 5140 667
% 
\special{pn 13}%
\special{pa 5140 598}%
\special{pa 5140 668}%
\special{fp}%
% LINE 1 0 3 0
% 12 2382 170 2060 331 2060 331 1833 484 1833 484 1680 640 1594 793 1527 950 1515 1268 1594 1424 1594 1424 1750 1577
% 
\special{pn 13}%
\special{pa 2382 170}%
\special{pa 2060 332}%
\special{fp}%
\special{pa 2060 332}%
\special{pa 1834 484}%
\special{fp}%
\special{pa 1834 484}%
\special{pa 1680 640}%
\special{fp}%
\special{pa 1594 794}%
\special{pa 1528 950}%
\special{fp}%
\special{pa 1516 1268}%
\special{pa 1594 1424}%
\special{fp}%
\special{pa 1594 1424}%
\special{pa 1750 1578}%
\special{fp}%
% LINE 1 0 3 0
% 4 2056 323 1045 323 1045 323 888 170
% 
\special{pn 13}%
\special{pa 2056 324}%
\special{pa 1046 324}%
\special{fp}%
\special{pa 1046 324}%
\special{pa 888 170}%
\special{fp}%
% STR 2 0 3 0
% 3 1280 1000 1280 1040 5 0
% $\mu_a$
\put(12.8000,-10.4000){\makebox(0,0){$\mu_a$}}%
% STR 2 0 3 0
% 3 1480 351 1480 390 5 0
% $\mu_1$
\put(14.8000,-3.9000){\makebox(0,0){$\mu_1$}}%
% STR 2 0 3 0
% 3 1260 1681 1260 1720 5 0
% (a)
\put(12.6000,-17.2000){\makebox(0,0){(a)}}%
% STR 2 0 3 0
% 3 4867 1741 4867 1780 2 0
% (b)
\put(48.6700,-17.8000){\makebox(0,0)[lb]{(b)}}%
% LINE 1 0 3 0
% 6 1045 327 1123 472 1123 472 1123 645 1123 789 1064 950
% 
\special{pn 13}%
\special{pa 1046 328}%
\special{pa 1124 472}%
\special{fp}%
\special{pa 1124 472}%
\special{pa 1124 646}%
\special{fp}%
\special{pa 1124 790}%
\special{pa 1064 950}%
\special{fp}%
% LINE 2 2 3 0
% 6 1116 653 1113 794 1673 645 1587 798 1517 959 1517 1277
% 
\special{pn 8}%
\special{pa 1116 654}%
\special{pa 1114 794}%
\special{dt 0.050}%
\special{pa 1674 646}%
\special{pa 1588 798}%
\special{dt 0.050}%
\special{pa 1518 960}%
\special{pa 1518 1278}%
\special{dt 0.050}%
% LINE 2 2 3 0
% 2 1064 954 927 1272
% 
\special{pn 8}%
\special{pa 1064 954}%
\special{pa 928 1272}%
\special{dt 0.050}%
% LINE 1 0 3 0
% 4 931 1272 814 1432 814 1432 637 1569
% 
\special{pn 13}%
\special{pa 932 1272}%
\special{pa 814 1432}%
\special{fp}%
\special{pa 814 1432}%
\special{pa 638 1570}%
\special{fp}%
% LINE 1 0 3 0
% 6 4733 330 4812 475 4812 475 4812 648 4812 793 4753 954
% 
\special{pn 13}%
\special{pa 4734 330}%
\special{pa 4812 476}%
\special{fp}%
\special{pa 4812 476}%
\special{pa 4812 648}%
\special{fp}%
\special{pa 4812 794}%
\special{pa 4754 954}%
\special{fp}%
% LINE 2 2 3 0
% 2 4750 954 4612 1271
% 
\special{pn 8}%
\special{pa 4750 954}%
\special{pa 4612 1272}%
\special{dt 0.050}%
% LINE 1 0 3 0
% 4 4620 1268 4502 1428 4502 1428 4326 1565
% 
\special{pn 13}%
\special{pa 4620 1268}%
\special{pa 4502 1428}%
\special{fp}%
\special{pa 4502 1428}%
\special{pa 4326 1566}%
\special{fp}%
% LINE 1 0 3 0
% 12 1115 468 1860 468 1127 645 1672 645 1123 793 1598 793 1061 954 1527 954 931 1268 1524 1268 818 1429 1594 1429
% 
\special{pn 13}%
\special{pa 1116 468}%
\special{pa 1860 468}%
\special{fp}%
\special{pa 1128 646}%
\special{pa 1672 646}%
\special{fp}%
\special{pa 1124 794}%
\special{pa 1598 794}%
\special{fp}%
\special{pa 1062 954}%
\special{pa 1528 954}%
\special{fp}%
\special{pa 932 1268}%
\special{pa 1524 1268}%
\special{fp}%
\special{pa 818 1430}%
\special{pa 1594 1430}%
\special{fp}%
% LINE 1 0 3 0
% 2 4733 327 4573 170
% 
\special{pn 13}%
\special{pa 4734 328}%
\special{pa 4574 170}%
\special{fp}%
% LINE 1 0 3 0
% 14 4738 315 5212 315 4808 467 5216 467 4812 640 5216 640 4812 797 5216 797 4753 954 5220 954 4616 1268 5220 1268 4510 1428 5220 1428
% 
\special{pn 13}%
\special{pa 4738 316}%
\special{pa 5212 316}%
\special{fp}%
\special{pa 4808 468}%
\special{pa 5216 468}%
\special{fp}%
\special{pa 4812 640}%
\special{pa 5216 640}%
\special{fp}%
\special{pa 4812 798}%
\special{pa 5216 798}%
\special{fp}%
\special{pa 4754 954}%
\special{pa 5220 954}%
\special{fp}%
\special{pa 4616 1268}%
\special{pa 5220 1268}%
\special{fp}%
\special{pa 4510 1428}%
\special{pa 5220 1428}%
\special{fp}%
% LINE 2 2 3 0
% 2 4816 644 4816 797
% 
\special{pn 8}%
\special{pa 4816 644}%
\special{pa 4816 798}%
\special{dt 0.050}%
% VECTOR 1 0 3 0
% 2 1064 954 1002 1107
% 
\special{pn 13}%
\special{pa 1064 954}%
\special{pa 1002 1108}%
\special{fp}%
\special{sh 1}%
\special{pa 1002 1108}%
\special{pa 1046 1054}%
\special{pa 1022 1058}%
\special{pa 1010 1038}%
\special{pa 1002 1108}%
\special{fp}%
% STR 2 0 3 0
% 3 1630 880 1630 920 2 0
% $(-a+m+2,1)$
\put(16.3000,-9.2000){\makebox(0,0)[lb]{$(-a+m+2,1)$}}%
% STR 2 0 3 0
% 3 2400 291 2400 330 2 0
% $(m+1,1)$
\put(24.0000,-3.3000){\makebox(0,0)[lb]{$(m+1,1)$}}%
% STR 2 0 3 0
% 3 2047 460 2047 499 2 0
% $(m,1)$
\put(20.4700,-4.9900){\makebox(0,0)[lb]{$(m,1)$}}%
% STR 2 0 3 0
% 3 986 1005 986 1044 3 0
% $(-a-1+N,-1)$
\put(9.8600,-10.4400){\makebox(0,0)[rb]{$(-a-1+N,-1)$}}%
% STR 2 0 3 0
% 3 1750 1551 1750 1590 2 0
% $(-N+m+1,1)$
\put(17.5000,-15.9000){\makebox(0,0)[lb]{$(-N+m+1,1)$}}%
% STR 2 0 3 0
% 3 1220 1471 1220 1510 5 0
% $\mu_N$
\put(12.2000,-15.1000){\makebox(0,0){$\mu_N$}}%
% STR 2 0 3 0
% 3 1500 201 1500 240 5 0
% $Q_{B_1}$
\put(15.0000,-2.4000){\makebox(0,0){$Q_{B_1}$}}%
% STR 2 0 3 0
% 3 1310 821 1310 860 5 0
% $Q_{B_a}$
\put(13.1000,-8.6000){\makebox(0,0){$Q_{B_a}$}}%
% STR 2 0 3 0
% 3 1230 1301 1230 1340 5 0
% $Q_{B_N}$
\put(12.3000,-13.4000){\makebox(0,0){$Q_{B_N}$}}%
% STR 2 0 3 0
% 3 1045 417 1045 456 3 0
% $Q_{F_1}$
\put(10.4500,-4.5600){\makebox(0,0)[rb]{$Q_{F_1}$}}%
% STR 2 0 3 0
% 3 833 1319 833 1358 3 0
% $Q_{F_{N-1}}$
\put(8.3300,-13.5800){\makebox(0,0)[rb]{$Q_{F_{N-1}}$}}%
% VECTOR 1 0 3 0
% 2 1524 954 1594 793
% 
\special{pn 13}%
\special{pa 1524 954}%
\special{pa 1594 794}%
\special{fp}%
\special{sh 1}%
\special{pa 1594 794}%
\special{pa 1550 846}%
\special{pa 1574 842}%
\special{pa 1586 862}%
\special{pa 1594 794}%
\special{fp}%
% STR 2 0 3 0
% 3 5462 278 5462 317 5 0
% $\mu_1$
\put(54.6200,-3.1700){\makebox(0,0){$\mu_1$}}%
% STR 2 0 3 0
% 3 5469 432 5469 471 5 0
% $\mu_2$
\put(54.6900,-4.7100){\makebox(0,0){$\mu_2$}}%
% STR 2 0 3 0
% 3 5490 1398 5490 1437 5 0
% $\mu_N$
\put(54.9000,-14.3700){\makebox(0,0){$\mu_N$}}%
% STR 2 0 3 0
% 3 4727 446 4727 485 3 0
% $Q_{F_1}$
\put(47.2700,-4.8500){\makebox(0,0)[rb]{$Q_{F_1}$}}%
% STR 2 0 3 0
% 3 4777 621 4777 660 3 0
% $Q_{F_2}$
\put(47.7700,-6.6000){\makebox(0,0)[rb]{$Q_{F_2}$}}%
% STR 2 0 3 0
% 3 4510 1363 4510 1402 3 0
% $Q_{F_N}$
\put(45.1000,-14.0200){\makebox(0,0)[rb]{$Q_{F_N}$}}%
% LINE 1 0 3 0
% 2 5140 282 5140 352
% 
\special{pn 13}%
\special{pa 5140 282}%
\special{pa 5140 352}%
\special{fp}%
% LINE 1 0 3 0
% 2 5112 282 5112 352
% 
\special{pn 13}%
\special{pa 5112 282}%
\special{pa 5112 352}%
\special{fp}%
% LINE 1 0 3 0
% 2 5112 429 5112 499
% 
\special{pn 13}%
\special{pa 5112 430}%
\special{pa 5112 500}%
\special{fp}%
% LINE 1 0 3 0
% 2 5140 429 5140 499
% 
\special{pn 13}%
\special{pa 5140 430}%
\special{pa 5140 500}%
\special{fp}%
% LINE 1 0 3 0
% 2 5112 758 5112 828
% 
\special{pn 13}%
\special{pa 5112 758}%
\special{pa 5112 828}%
\special{fp}%
% LINE 1 0 3 0
% 2 5140 758 5140 828
% 
\special{pn 13}%
\special{pa 5140 758}%
\special{pa 5140 828}%
\special{fp}%
% LINE 1 0 3 0
% 2 5112 919 5112 989
% 
\special{pn 13}%
\special{pa 5112 920}%
\special{pa 5112 990}%
\special{fp}%
% LINE 1 0 3 0
% 2 5140 919 5140 989
% 
\special{pn 13}%
\special{pa 5140 920}%
\special{pa 5140 990}%
\special{fp}%
% LINE 1 0 3 0
% 2 5112 1234 5112 1304
% 
\special{pn 13}%
\special{pa 5112 1234}%
\special{pa 5112 1304}%
\special{fp}%
% LINE 1 0 3 0
% 2 5140 1234 5140 1304
% 
\special{pn 13}%
\special{pa 5140 1234}%
\special{pa 5140 1304}%
\special{fp}%
% LINE 1 0 3 0
% 2 5112 1402 5112 1472
% 
\special{pn 13}%
\special{pa 5112 1402}%
\special{pa 5112 1472}%
\special{fp}%
% LINE 1 0 3 0
% 2 5140 1402 5140 1472
% 
\special{pn 13}%
\special{pa 5140 1402}%
\special{pa 5140 1472}%
\special{fp}%
\end{picture}%
\end{center} 
\caption{(a)The toric diagram of $SU(N)$ geometry (b)The building block of $SU(N)$ geometry, and refined vertex on this geometry implies $\mathop K\nolimits_{\mathop \mu \nolimits_1  \cdots \mathop \mu \nolimits_N } \left( {\mathop Q\nolimits_{F,1} , \cdots ,\mathop Q\nolimits_{F,N - 1} } \right)$}
\label{suN}
\end{figure}

The toric diagram of $SU(N)$ geometry which give rise to the $\cal N$$=2$ $SU(N)$ super Yang-Mills are shown in Fig.\ref{suN}(a). The parallel edges corresponding to the base ${\mathbb P}^{1} $ are the preferred directions of \cite{Iqbal:2007ii}. For fixed $N$, there are $N+1$ inequivalent geometries ($m = 0 \cdots N $) which give $SU(N)$ super Yang-Mills. The number $m$ is the Chern-Simons coefficient of the 5-dimensiomal theory in the gauge theory side.
%%%%%%%%%%%%%%%%%%%%%%%%%%%%%%%%%%%%%%%%%%%%%%%%%%%%%%%%%%%%

Let us start with the computation of the subdiagram Fig.\ref{suN}(b). For the reason which we discuss later, we modify slightly the framing factor proposed in \cite{Iqbal:2007ii} as follows
\begin{align}
\label{modification}
\mathop f\nolimits_\mu  (t,q) = \mathop {( - 1)}\nolimits^{\left| \mu  \right|} \mathop t\nolimits^{\frac{{|| {\mu ^t } ||^2 }}{2}} \mathop q\nolimits^{ - \frac{{\left\| \mu  \right\|^2 }}{2}} 
 = \mathop {( - 1)}\nolimits^{\left| \mu  \right|} \mathop {\left( {\frac{t}{q}} \right)}\nolimits^{\frac{{\mathop {|| {\mu ^t }||}\nolimits^2 }}{2}} \mathop q\nolimits^{ - \frac{{\mathop \kappa \nolimits_\mu  }}{2}} 
\end{align}
Using the refined vertex, we can express the subdiagram as
\begin{align}
\mathop K\nolimits_{\mathop \mu \nolimits_1  \cdots \mathop \mu \nolimits_N } \left( {\mathop Q\nolimits_{F,1} , \cdots ,\mathop Q\nolimits_{F,N - 1} } \right) &= \sum\limits_{\mathop \lambda \nolimits_1  \cdots \mathop \lambda \nolimits_{N - 1} } {\prod\limits_{a = 1}^N {\left( { - \mathop Q\nolimits_{F,a} } \right)^{|| {\mathop \lambda \nolimits_a } ||} \mathop f\nolimits_{\mathop \lambda \nolimits_a } (t,q)\mathop C\nolimits_{\mathop \lambda \nolimits_{a - 1} ^t \mathop \lambda \nolimits_a \mathop \mu \nolimits_a } (t,q)} }  \nonumber\\
 &= \sum\limits_{\mathop \lambda \nolimits_1  \cdots \mathop \lambda \nolimits_{N - 1} } {\prod\limits_{a = 1}^N {\sum\limits_{\mathop \eta \nolimits_1  \cdots \mathop \eta \nolimits_N } {\mathop Q\nolimits_{F,a} ^{\left| {\mathop \lambda \nolimits_a } \right|} \mathop {\left( {\frac{t}{q}} \right)}\nolimits^{\frac{{|| {\mathop \lambda \nolimits_a ^t } ||^2 }}{2}} \mathop q\nolimits^{ - \frac{{\mathop \kappa \nolimits_{\mathop \lambda \nolimits_a } }}{2}} \mathop {\left( {\frac{q}{t}} \right)}\nolimits^{\frac{{\left\| {\mathop \lambda \nolimits_a } \right\|^2  + \left\| {\mathop \mu \nolimits_a } \right\|^2 }}{2}} } } } \nonumber\\
&\times \mathop t\nolimits^{\frac{{\mathop \kappa \nolimits_{\mathop \lambda \nolimits_a } }}{2}} \mathop P\nolimits_{\mathop \mu \nolimits_a ^t } (\mathop t\nolimits^{ - \rho } ;q,t)\mathop {\left( {\frac{q}{t}} \right)}\nolimits^{\frac{{\left| {\mathop \eta \nolimits_a } \right| + \left| {\mathop \lambda \nolimits_{a - 1} } \right| - \left| {\mathop \lambda \nolimits_a } \right|}}{2}} \cr 
&\times \mathop s\nolimits_{\mathop \lambda \nolimits_{a - 1} /\mathop \eta \nolimits_a } (\mathop t\nolimits^{ - \rho } \mathop q\nolimits^{ - \mathop \mu \nolimits_a } )\mathop s\nolimits_{\mathop \lambda \nolimits_a /\mathop \eta \nolimits_a } (\mathop t\nolimits^{ - \mathop {\mathop \mu \nolimits_a }\nolimits^t } \mathop q\nolimits^{ - \rho } )
\end{align}
Notice $\left\| \mu  \right\|^2  - \left\| {\mu ^t } \right\|^2  = \mathop \kappa \nolimits_\mu  $ and $\mathop \lambda \nolimits_0  = \mathop \lambda \nolimits_N  = \phi $. Simplifing the summation, we get
\begin{align}
\mathop K\nolimits_{\mathop \mu \nolimits_1  \cdots \mathop \mu \nolimits_N } \left( {\mathop Q\nolimits_{F,1} , \cdots ,\mathop Q\nolimits_{F,N - 1} } \right) &= \prod\limits_{a = 1}^N {\left[ {\mathop q\nolimits^{\frac{{\left\| {\mathop \mu \nolimits_a } \right\|^2 }}{2}} \mathop {\tilde Z}\nolimits_{\mu_a } (t,q)} \right]} 
 \cr 
& \times \sum\limits_{\mathop \lambda \nolimits_1  \cdots \mathop \lambda \nolimits_{N - 1} } {\sum\limits_{\mathop \eta \nolimits_1  \cdots \mathop \eta \nolimits_N } {\prod\limits_{a = 1}^N {\mathop Q\nolimits_{F,a} ^{\left| {\mathop \lambda \nolimits_a } \right|} \mathop {\left( {\frac{q}{t}} \right)}\nolimits^{\frac{{\left| {\mathop \eta \nolimits_a } \right|}}{2}} \mathop s\nolimits_{\mathop \lambda \nolimits_{a - 1} /\mathop \eta \nolimits_a } (\mathop t\nolimits^{ - \rho } \mathop q\nolimits^{ - \mathop \mu \nolimits_a } )\mathop s\nolimits_{\mathop \lambda \nolimits_a /\mathop \eta \nolimits_a } (\mathop t\nolimits^{ - \mathop {\mathop \mu \nolimits_a }\nolimits^t } \mathop q\nolimits^{ - \rho } )} } } \nonumber
\end{align}
The sum involved in the above subdiagram becomes
\begin{align}
&\sum\limits_{\mathop \lambda \nolimits_1  \cdots \mathop \lambda \nolimits_{N - 1} } {\sum\limits_{\mathop \eta \nolimits_1  \cdots \mathop \eta \nolimits_N } {\prod\limits_{a = 1}^N {\left( {\sqrt {\frac{q}{t}} \mathop Q\nolimits_{F,a} } \right)^{\left| {\mathop \lambda \nolimits_a } \right|} \mathop s\nolimits_{\mathop \lambda \nolimits_{a - 1} /\mathop \eta \nolimits_a } (\mathop t\nolimits^{ - \rho } \mathop q\nolimits^{ - \mathop \mu \nolimits_a  - \frac{1}{2}} )\mathop s\nolimits_{\mathop \lambda \nolimits_a /\mathop \eta \nolimits_a } (\mathop t\nolimits^{ - \mathop {\mathop \mu \nolimits_a }\nolimits^t  + \frac{1}{2}} \mathop q\nolimits^{ - \rho } )} } } \cr
 &= \sum\limits_{\mathop \lambda \nolimits_1  \cdots \mathop \lambda \nolimits_{N - 1} } {\sum\limits_{\mathop \rho \nolimits_1  \cdots \mathop \rho \nolimits_{N-1} } {\prod\limits_{a = 1}^{N-1} {\left( {\sqrt {\frac{q}{t}} \mathop Q\nolimits_{F,a} } \right)^{\left| {\mathop \lambda \nolimits_a } \right|} \mathop s\nolimits_{\mathop \lambda \nolimits_a /\mathop \rho \nolimits_{a - 1} } (\mathop t\nolimits^{ - \mathop {\mathop \mu \nolimits_a }\nolimits^t  + \frac{1}{2}} \mathop q\nolimits^{ - \rho } )\mathop s\nolimits_{\mathop \lambda \nolimits_a /\mathop \rho \nolimits_a } (\mathop t\nolimits^{ - \rho } \mathop q\nolimits^{ - \mathop \mu \nolimits_{a + 1}  - \frac{1}{2}} )} } } \nonumber
\end{align}
Notice taht $\mathop \rho \nolimits_0  = \mathop \rho \nolimits_{N - 1}  = \phi $. We can take the summation over Young diagrams by Lemma 3.1 of \cite{Zhou:2003}, or using the vertex on a strip \cite{Iqbal:2004ne} as we will disscuss in the next subsection. Then we get
\begin{align}
\label{pureK}
\mathop K\nolimits_{\mathop \mu \nolimits_1  \cdots \mathop \mu \nolimits_N } \left( {\mathop Q\nolimits_{F,a} } \right) &= \prod\limits_{a = 1}^N {\left[ {\mathop q\nolimits^{\frac{{\left\| {\mathop \mu \nolimits_a } \right\|^2 }}{2}} \mathop {\tilde Z}\nolimits_{\mathop \mu \nolimits_a } (t,q)} \right]} 
 \cr
& \times \prod\limits_{1 \le a < b \le N} {\prod\limits_{i,j = 1}^\infty  {\frac{1}{{1 - \mathop Q\nolimits_{ab} \mathop t\nolimits^{ - \mathop \mu \nolimits_{ai} ^t  + j} \mathop q\nolimits^{ - \mathop \mu \nolimits_{bj}  + i - 1} }}} } 
\end{align}
where $\mathop Q\nolimits_{ab}  \equiv \prod\limits_{l = a}^{b - 1} {\left( {\sqrt {\frac{q}{t}} \mathop Q\nolimits_{F,l} } \right)}  \equiv \prod\limits_{l = a}^{b - 1} {\mathop {\tilde Q}\nolimits_{F,l} } 
$. 

Let us glue these subdiagrams. The framing factors are given by
$\mathop n\nolimits_a  = \left( {a - m - 2, - 1} \right) \wedge \left( {a - N + 1,1} \right) = - ( N + m - 2a + 1 )$ as Fig.\ref{suN}.  Then, the A-model amplitude is
\begin{align}
\label{pureZ}
\mathop Z\nolimits^{A - model,SU(N)} \left( {\mathop Q\nolimits_B ,\mathop Q\nolimits_{F,a} } \right) &= \sum\limits_{\mathop \mu \nolimits_1  \cdots \mathop \mu \nolimits_N } {\prod\limits_{a = 1}^N {\left[ {\mathop Q\nolimits_{B,a} ^{\left| {\mathop \mu \nolimits_a } \right|} \mathop f\nolimits_{\mathop \mu \nolimits_a } \left( {t,q} \right)^{\mathop n\nolimits_a } } \right]} \mathop K\nolimits_{\mathop \mu \nolimits_1  \cdots \mathop \mu \nolimits_N } \left( {\mathop Q\nolimits_{F,a}, t,q} \right)\mathop K\nolimits_{\mathop \mu \nolimits_N ^t  \cdots \mathop \mu \nolimits_1 ^t } \left( {\mathop Q\nolimits_{F,a} ,q,t} \right)} \cr
 &= \mathop Z\nolimits_{pert.}^{A - model,SU(N)} \left( {\mathop Q\nolimits_{F,a} } \right)\mathop Z\nolimits_{inst}^{A - model,SU(N)} \left( {\mathop Q\nolimits_B ,\mathop Q\nolimits_{F,a} } \right)
\end{align}
The perturbative part of the partition function is given by \cite{Nekrasov:2002qd}
\begin{eqnarray*}
\mathop Z\nolimits_{pert.}^{A - model,SU(N)} \left( {\mathop Q\nolimits_{F,a} } \right) \equiv \mathop K\nolimits_{\phi  \cdots \phi } \left( {\mathop Q\nolimits_{F,a} } \right)^2 
\end{eqnarray*}
By substituting (\ref{pureK}) into (\ref{pureZ}), we obtain
\begin{align}
\mathop Z\nolimits_{inst.}^{A - model,SU(N)} \left( {\mathop Q\nolimits_B ,\mathop Q\nolimits_{F,a} } \right) &= \sum\limits_{\mathop \mu \nolimits_1  \cdots \mathop \mu \nolimits_N } {\prod\limits_{a = 1}^N {\left[ {\mathop Q\nolimits_{B,a} ^{\left| {\mathop \mu \nolimits_a } \right|} \mathop f\nolimits_{\mathop \mu \nolimits_a } \left( {t,q} \right)^{\mathop n\nolimits_a } \mathop q\nolimits^{\frac{{\left\| {\mathop \mu \nolimits_a } \right\|^2 }}{2}} \mathop t\nolimits^{\frac{{|| {\mathop \mu \nolimits_a^t }||^2 }}{2}} \mathop {\tilde Z}\nolimits_{\mathop \mu \nolimits_a } (t,q)\mathop {\tilde Z}\nolimits_{\mathop \mu \nolimits_a^t } (q,t)} \right]} } \cr
& \hspace{-1.5cm}\times \prod\limits_{1 \le a < b \le N} {\prod\limits_{i,j = 1}^\infty  {\frac{{1 - \mathop Q\nolimits_{ab} \mathop t\nolimits^j \mathop q\nolimits^{i - 1} }}{{1 - \mathop Q\nolimits_{ab} \mathop t\nolimits^{ - \mathop \mu \nolimits_{ai} ^t  + j} \mathop q\nolimits^{ - \mathop \mu \nolimits_{bj}  + i - 1} }}} \frac{{1 - \mathop Q\nolimits_{ab} \mathop t\nolimits^{j - 1} \mathop q\nolimits^i }}{{1 - \mathop Q\nolimits_{ab} \mathop t\nolimits^{ - \mathop \mu \nolimits_{ai} ^t  + j - 1} \mathop q\nolimits^{ - \mathop \mu \nolimits_{bj}  + i} }}} 
\end{align}
As we show in the next subsection, the partition function is identical with that of Nekrasov.

%%%%%%%%%%%%%%%%%%%%%%%%%%%%%%%%%%%%%%%%%%%%%%%%%%%%%%%%%%%%%%%
\subsubsection{Identification with Nekrasov's Partition Functions}
%%%%%%%%%%%%%%%%%%%%%%%%%%%%%%%%%%%%%%%%%%%%%%%%%%%%%%%%%%%%%%%

In this subsection, we show that the refined A-model amplitude agrees with the K-theoretic version of the Nekrasov's partition function:
\begin{align}
\mathop Z\nolimits_{inst.}^{A - model,SU(N)} \left( {\mathop Q\nolimits_B ,\mathop Q\nolimits_{F,a} } \right) = \mathop Z\nolimits_{inst.}^{Nek,SU(N)} \left( {\hat Q,\mathop Q\nolimits_{ab} } \right)
\end{align}
Recall that the K-theoretic version of the Nekrasov's partition functions with a Chern-Simons term is given by \cite{Tachikawa:2004ur}\cite{Gottsche:2007}
\begin{align}
\label{nek:withcs}
\mathop Z\nolimits_{inst.}^{Nek.SU(N),m} (\hat Q,\mathop Q\nolimits_{ab} ) = \sum\limits_{\mathord{\buildrel{\lower3pt\hbox{$\scriptscriptstyle\rightharpoonup$}} 
\over \mu } } {\frac{{\mathop {\hat Q}\nolimits^{\left| {\vec \mu } \right|} }}{{\prod\limits_{a , b} {\mathop N\nolimits_{ab}^{\vec \mu } (t,q,\mathop Q\nolimits_{ab} )} }}\mathop {\left( {\frac{q}{t}} \right)}\nolimits^{\frac{N}{2}\left| {\vec \mu } \right|} \prod\limits_{a = 1}^N {\mathop e\nolimits_a ^{m\left| {\mathop \mu \nolimits_a } \right|} \mathop t\nolimits^{ - m\frac{{|| {\mathop \mu \nolimits_a^t }||^2 }}{2}} \mathop q\nolimits^{m\frac{{\left\| {\mathop \mu \nolimits_a } \right\|^2 }}{2}} } } 
\end{align}
Note that $\mathop Q\nolimits_{ab}  = \mathop e\nolimits_a \mathop e\nolimits_b^{ - 1} $. 

%%%%%%%%%%%%%%%%%%%%%%%%%%%%%%%%%%%%%%%%%%%%%%%%%%
First, let us rewrite the character part $
\prod {\mathop N\nolimits_{ab}^{\vec \mu } } 
$. The identity 
\begin{align}
\label{id:3-1}
\sum\limits_{i,j = 1}^\infty  {\mathop q\nolimits^{\mathop \mu \nolimits_i  - j + 1} \mathop t\nolimits^{\mathop \nu \nolimits_j  - i} }  = \sum\limits_{i,j = 1}^\infty  {\mathop q\nolimits^{ - \mathop \nu \nolimits_j^t  + i} \mathop t\nolimits^{ - \mathop \mu \nolimits_i^t  + j - 1} } 
\end{align}
follows from
$\left( {t - 1} \right)\sum\limits_{i = 1}^\infty  {\mathop q\nolimits^{\mathop \mu \nolimits_i } \mathop t\nolimits^{ - i} }  = \left( {\mathop q\nolimits^{ - 1}  - 1} \right)\sum\limits_{i = 1}^\infty  {\mathop t\nolimits^{ - \mathop \mu \nolimits_i^t } \mathop q\nolimits^i } 
$ for $t,q \ne 1$ \cite{Awata:2005fa}. It is easy to prove the following formula using (\ref{id:3-1}) (take the logarithm of the equation(\ref{id:3-2}))
\begin{align}
\label{id:3-2}
\prod\limits_{i,j = 1}^\infty  {\left( {1 - Q\mathop t\nolimits^{ - \mathop \mu \nolimits_j ^t  + i} \mathop q\nolimits^{ - \mathop \nu \nolimits_i  + j - 1} } \right)}  = \prod\limits_{i,j = 1}^\infty  {\left( {1 - Q\mathop q\nolimits^{\mathop \mu \nolimits_i  - j} \mathop t\nolimits^{\mathop \nu \nolimits_j ^t  - i + 1} } \right)} 
\end{align}
The character part of the Nekrasov's patririon function is given by
\begin{align}
\frac{1}{{\mathop N\nolimits_{12}^{\vec \mu } \left( {t,q,Q} \right)}} &\equiv \prod\limits_{\left( {i,j} \right) \in \mu } {\frac{1}{{1 - Q\mathop t\nolimits^{  \mathop \nu \nolimits_j ^t  - i} \mathop q\nolimits^{  \mathop \mu \nolimits_i  - j + 1} }}} \prod\limits_{\left( {i,j} \right) \in \nu } {\frac{1}{{1 - Q\mathop t\nolimits^{ - \mathop \mu \nolimits_j ^t  + i - 1} \mathop q\nolimits^{ - \mathop \nu \nolimits_i  + j} }}} \nonumber\\
 &= \prod\limits_{i,j = 1}^\infty  {\frac{{1 - Q\mathop t\nolimits^{j-1} \mathop q\nolimits^{i } }}{{1 - Q\mathop t\nolimits^{ -\mathop \mu \nolimits_j ^t  + i - 1} \mathop q\nolimits^{ - \mathop \nu \nolimits_i  + j} }}} 
\end{align}
where $\mathop \mu \nolimits_1  = \mu ,\mathop \mu \nolimits_2  = \nu ,\mathop Q\nolimits_{12}  = Q$. By using (\ref{id:3-2}), we have
\begin{align}
\prod\limits_{i,j = 1}^\infty  {\frac{{1 - Q\mathop t\nolimits^j \mathop q\nolimits^{i - 1} }}{{1 - Q\mathop t\nolimits^{ - \mathop \mu \nolimits_j ^t  + i} \mathop q\nolimits^{ - \mathop \nu \nolimits_i  + j - 1} }}}  &= \prod\limits_{i,j = 1}^\infty  {\frac{{1 - Q\mathop t\nolimits^j \mathop q\nolimits^{i - 1} }}{{1 - Q\mathop q\nolimits^{\mathop \mu \nolimits_i  - j} \mathop t\nolimits^{\mathop \nu \nolimits_j ^t  - i + 1} }}} \nonumber\\
&= \prod\limits_{\left( {i,j} \right) \in \nu } {\frac{1}{{1 - Q\mathop t\nolimits^{ - \mathop \mu \nolimits_j ^t  + i} \mathop q\nolimits^{ - \mathop \nu \nolimits_i  + j - 1} }}} \prod\limits_{\left( {i,j} \right) \in \mu } {\frac{1}{{1 - Q\mathop t\nolimits^{  \mathop \nu \nolimits_j ^t  - i + 1} \mathop q\nolimits^{ \mathop \mu \nolimits_i  - j} }}} \nonumber\\
 &= \mathop {(-Q)}\nolimits^{ - \left| \mu  \right| - \left| \nu  \right|} \mathop t\nolimits^{\sum\limits_{\left( {i,j} \right) \in \nu } {\left( {\mathop \mu \nolimits_j ^t  - i} \right)}  - \sum\limits_{\left( {i,j} \right) \in \mu } {\left( {\mathop \nu \nolimits_j ^t  - i + 1} \right)} } \mathop q\nolimits^{\sum\limits_{\left( {i,j} \right) \in \nu } {\left( {\mathop \nu \nolimits_i  - j + 1} \right)}  - \sum\limits_{\left( {i,j} \right) \in \mu } {\left( {\mathop \mu \nolimits_i  - j} \right)} } \nonumber\\
& \times \frac{1}{{\mathop N\nolimits_{21}^{\vec \mu } \left( {t,q,Q^{ - 1} } \right)}}
\end{align}
The factors appear in the above equation become
\begin{align}
\sum\limits_{(i,j) \in \nu } {\mathop \mu \nolimits_j^t }  = \sum\limits_{j = 1}^{\mathop \nu \nolimits_1 } {\sum\limits_{i = 1}^{\mathop \nu \nolimits_j^t } {\mathop \mu \nolimits_j^t } }  = \sum\limits_{j = 1}^{\min (\mathop \mu \nolimits_1 ,\mathop \nu \nolimits_1 )} {\mathop \mu \nolimits_j^t \mathop \nu \nolimits_j^t }  = \sum\limits_{(i,j) \in \mu } {\mathop \nu \nolimits_j^t } \nonumber
\end{align}
\begin{align}
\sum\limits_{\left( {i,j} \right) \in \mu } {\left( {\mathop \mu \nolimits_i  - j} \right)}  = \sum\limits_{i = 1}^{d(\mu )} {\left[ {\left( {\mathop \mu \nolimits_i  - 1} \right) +  \cdots  + \left( {\mathop \mu \nolimits_i  - \mathop \mu \nolimits_i } \right)} \right]} 
=  \frac{{\left\| \mu  \right\|^2 }}{2} - \frac{{\left| \mu  \right|}}{2}\nonumber
\end{align}
Hence we obtain
\begin{align}
\frac{1}{{\mathop N\nolimits_{12}^{\vec \mu } \left( {t,q,Q} \right)\mathop N\nolimits_{21}^{\vec \mu } \left( {t,q,Q^{ - 1} } \right)}} &= \mathop {(-Q)}\nolimits^{\left| \mu  \right| + \left| \nu  \right|} \left( {\frac{q}{t}} \right)^{ - \frac{{\left| \mu  \right|}}{2} - \frac{{\left| \nu  \right|}}{2} + \frac{{|| {\mu ^t } ||^2 }}{2} - \frac{{|| {\nu ^t }  ||^2 }}{2}} \mathop q\nolimits^{\frac{{\mathop \kappa \nolimits_\mu  }}{2} - \frac{{\mathop \kappa \nolimits_\nu  }}{2}} \cr
& \times \prod\limits_{i,j = 1}^\infty  {\frac{{1 - \mathop Q\nolimits_{12} \mathop t\nolimits^{i - 1} \mathop q\nolimits^j }}{{1 - \mathop Q\nolimits_{12} \mathop t\nolimits^{ - \mathop \mu \nolimits_j ^t  + i - 1} \mathop q\nolimits^{ - \mathop \nu \nolimits_i  + j} }}\frac{{1 - \mathop Q\nolimits_{12} \mathop t\nolimits^i \mathop q\nolimits^{j - 1} }}{{1 - \mathop Q\nolimits_{12} \mathop t\nolimits^{ - \mathop \mu \nolimits_j ^t  + i} \mathop q\nolimits^{ - \mathop \nu \nolimits_i  + j - 1} }}} 
\end{align}
It is easy to show 
\begin{align}
\frac{1}{{\mathop N\nolimits_{aa}^{\vec \mu } \left( {t,q,\mathop Q\nolimits_{aa}  = 1} \right)}} = (-1)^{|\mu_a|} \mathop {\left( {\frac{t}{q}} \right)}\nolimits^{\frac{{\left| {\mathop \mu \nolimits_a } \right|}}{2}} \mathop t\nolimits^{\frac{{|| {\mathop \mu \nolimits_a^t } ||^2 }}{2}} \mathop q\nolimits^{\frac{{\left\| {\mathop \mu \nolimits_a } \right\|^2 }}{2}} \mathop {\tilde Z}\nolimits_{\mathop \mu \nolimits_a } \left( {t,q} \right)\mathop {\tilde Z}\nolimits_{\mathop \mu \nolimits_a^t } \left( {q,t} \right)
\end{align}
By combining above identities, we can rewrite the Nekrasov's partition function as follows
\begin{align}
\label{nek:rewritten}
&\sum\limits_{\mathord{\buildrel{\lower3pt\hbox{$\scriptscriptstyle\rightharpoonup$}} 
\over \mu } } {\frac{{\mathop {\hat Q}\nolimits^{\left| {\vec \mu } \right|} }}{{\prod\limits_{a < b} {\mathop N\nolimits_{ab}^{\vec \mu } (t,q,\mathop Q\nolimits_{ab} )} }}\mathop {\left( {\frac{q}{t}} \right)}\nolimits^{\frac{N}{2}\left| {\vec \mu } \right|} \prod\limits_{a = 1}^N {\mathop e\nolimits_a ^{m\left| {\mathop \mu \nolimits_a } \right|} \mathop t\nolimits^{ - m\frac{{|| {\mathop \mu \nolimits_a^t } ||^2 }}{2}} \mathop q\nolimits^{m\frac{{\left\| {\mathop \mu \nolimits_a } \right\|^2 }}{2}} } } \nonumber\\
&= \sum\limits_{\vec \mu } (-1)^{N\left| {\vec \mu } \right|} 
{\mathop {\hat Q}\nolimits^{\left| {\vec \mu } \right|} \prod\limits_{a < b} {\left[ {\mathop Q\nolimits_{ab} ^{\left| {\mathop \mu \nolimits_a } \right| + \left| {\mathop \mu \nolimits_b } \right|} } \right]} \prod\limits_{a = 1}^N {[\mathop e\nolimits_a ^{m\left| {\mathop \mu \nolimits_a } \right|} \mathop {\left( {\frac{q}{t}} \right)}\nolimits^{(N + m - 2a + 1)\frac{{|| {\mathop \mu \nolimits_a^t } ||^2 }}{2}} \mathop q\nolimits^{(N + m - 2a + 1)\frac{{\mathop \kappa \nolimits_{\mathop \mu \nolimits_a } }}{2}} } } 
\nonumber\\
& \times \mathop t\nolimits^{\frac{{|| {\mathop \mu \nolimits_a^t } ||^2 }}{2}} \mathop q\nolimits^{\frac{{\left\| {\mathop \mu \nolimits_a } \right\|^2 }}{2}} \mathop {\tilde Z}\nolimits_{\mathop \mu \nolimits_a } (t,q)\mathop {\tilde Z}\nolimits_{\mathop \mu \nolimits_a^t } (q,t)] \nonumber\\ 
 & \times \prod\limits_{a < b} {\prod\limits_{i,j = 1}^\infty  {\frac{{1 - \mathop Q\nolimits_{ab} \mathop t\nolimits^{i - 1} \mathop q\nolimits^j }}{{1 - \mathop Q\nolimits_{ab} \mathop t\nolimits^{ - \mathop \mu \nolimits_{aj} ^t  + i - 1} \mathop q\nolimits^{ - \mathop \mu \nolimits_{bi}  + j} }}\frac{{1 - \mathop Q\nolimits_{ab} \mathop t\nolimits^i \mathop q\nolimits^{j - 1} }}{{1 - \mathop Q\nolimits_{ab} \mathop t\nolimits^{ - \mathop \mu \nolimits_{aj} ^t  + i} \mathop q\nolimits^{ - \mathop \mu \nolimits_{bi}  + j - 1} }}} } 
\end{align}

Next, let us rewrite the remainder ${\mathop {\hat Q}\nolimits^{\left| {\vec \mu } \right|} \prod\limits_{a < b} {\mathop Q\nolimits_{ab} ^{\left| {\mathop \mu \nolimits_a } \right| + \left| {\mathop \mu \nolimits_b } \right|} } \prod\limits_{a = 1}^N {\mathop e\nolimits_a ^{m\left| {\mathop \mu \nolimits_a } \right|} } }$. We shall rewrite it in terms of the K\"{a}hler parameters of the base  and the fiber $\mathop {\mathbb P}\nolimits^1 $'s by showing the following identity
\begin{align}
\label{id:3-3}
\mathop {C \mathop Q\nolimits_B }\nolimits^{\left| {\vec \mu } \right|} \prod\limits_{a < b} {\mathop Q\nolimits_{ab} ^{\left| {\mathop \mu \nolimits_a } \right| + \left| {\mathop \mu \nolimits_b } \right|} } \prod\limits_{a = 1}^N {\mathop e\nolimits_a ^{m\left| {\mathop \mu \nolimits_a } \right|} }  = \prod\limits_{a = 1}^N {\mathop Q\nolimits_{B,a} ^{\left| {\mathop \mu \nolimits_a } \right|} } \end{align}
We prove this identity in the case of $N=$odd and $m=$even for example. It is easy to generarize this proof. First we use the results of \cite{Iqbal:2003zz}, that is, ${\mathop Q\nolimits_{B,a} }$ are given by the base and the fiber K\"{a}hler parameters and they satisfy
\begin{align}
\label{formula01}
\prod\limits_{i = 1}^N {\mathop Q\nolimits_{B,a} ^{\left| {\mathop \mu \nolimits_a } \right|} }  = \mathop {\mathop Q\nolimits_B }\nolimits^{\left| {\vec \mu } \right|} \prod\limits_{a = 1}^{\left[ {\frac{{N + m - 1}}{2}} \right]} {\mathop {\tilde Q}\nolimits_{F,a} ^{(N + m - 2a)(\left| {\mathop \mu \nolimits_1 } \right| +  \cdots  + \left| {\mathop \mu \nolimits_a } \right|)} } \prod\limits_{a = \left[ {\frac{{N + m}}{2} + 1} \right]}^{N - 1} {\mathop {\tilde Q}\nolimits_{F,a} ^{ - (N + m - 2a)(\left| {\mathop \mu \nolimits_{a + 1} } \right| +  \cdots  + \left| {\mathop \mu \nolimits_N } \right|)} } 
\end{align}
Here we modify ${\mathop Q\nolimits_{F,a} }$ to ${\mathop {\tilde Q}\nolimits_{F,a} }$ in the case of the refined partition function. 

On the one hand we can obtain the following identity after some algebra
\begin{align}
\label{formula1}
\prod\limits_{a < b} {\mathop Q\nolimits_{ab} ^{\left| {\mathop \mu \nolimits_a } \right| + \left| {\mathop \mu \nolimits_b } \right|} }  &=\left ( {\prod\limits_{a = 1}^{\frac{{N - 1}}{2}} {\mathop {\tilde Q}\nolimits_{F,a} ^a } \prod\limits_{a = \frac{{N + 1}}{2}}^{N - 1} {\mathop {\tilde Q}\nolimits_{F,a} ^{N - a} } } \right)^{\left| {\vec \mu } \right|} \prod\limits_{a = 1}^{\frac{{N - 1}}{2}} {\left( {\prod\limits_{b = a}^{\frac{{N - 1}}{2}} {\mathop {\tilde Q}\nolimits_{F,b} ^{(N - 2b)} } } \right)^{\left| {\mathop \mu \nolimits_a } \right|} } \prod\limits_{a = \frac{{N + 1}}{2}+1}^{N - 1} {\left( {\prod\limits_{b = \frac{{N + 1}}{2}}^{a - 1} {\mathop {\tilde Q}\nolimits_{F,b} ^{(2b - N)} } } \right)^{\left| {\mathop \mu \nolimits_a } \right|} } \nonumber\\
&= \left( {\prod\limits_{a = 1}^{\frac{{N - 1}}{2}} {\mathop {\tilde Q}\nolimits_{F,a} ^a } \prod\limits_{a = \frac{{N + 1}}{2}}^{N - 1} {\mathop {\tilde Q}\nolimits_{F,a} ^{N - a} } } \right)^{\left| {\mathord{\buildrel{\lower3pt\hbox{$\scriptscriptstyle\rightharpoonup$}} 
\over \mu } } \right|} \prod\limits_{a = 1}^{\frac{{N - 1}}{2}} {\mathop {\tilde Q}\nolimits_{F,a} ^{(N - 2a)(\left| {\mathop \mu \nolimits_1 } \right| +  \cdots  + \left| {\mathop \mu \nolimits_a } \right|)} } \prod\limits_{a = \frac{{N + 1}}{2}}^{N - 1} {\mathop {\tilde Q}\nolimits_{F,a} ^{(2a - N)(\left| {\mathop \mu \nolimits_{a + 1} } \right| +  \cdots  + \left| {\mathop \mu \nolimits_N } \right|)} }  
\end{align}
Here we use $\mathop Q\nolimits_{ab}  = \prod\limits_{l = a}^{b - 1} {\mathop {\tilde Q}\nolimits_{F,l} } $. Using $\mathop {\tilde Q}\nolimits_{F,a}  = \mathop e\nolimits_a \mathop {\mathop e\nolimits_{a + 1} }\nolimits^{ - 1} $, we can also show
\begin{align}
\label{formula2}
&\prod\limits_{a = 1}^{\frac{{N - 1}}{2}} {\mathop {\tilde Q}\nolimits_{F,a} ^{m(\left| {\mathop \mu \nolimits_1 } \right| +  \cdots  + \left| {\mathop \mu \nolimits_a } \right|)} } \prod\limits_{a = \frac{{N + 1}}{2}}^{N - 1} {\mathop {\tilde Q}\nolimits_{F,a} ^{ - m(\left| {\mathop \mu \nolimits_{a + 1} } \right| +  \cdots  + \left| {\mathop \mu \nolimits_N } \right|)} }    \nonumber\\
&\times
\prod\limits_{a = \frac{{N + 1}}{2}}^{\frac{{N + m - 1}}{2}} {\mathop {\tilde Q}\nolimits_{F,a} ^{(N + m - 2a)(\left| {\mathop \mu \nolimits_1 } \right| +  \cdots  + \left| {\mathop \mu \nolimits_a } \right|)} } \prod\limits_{a = \frac{{N + 1}}{2}}^{\frac{{N + m - 1}}{2}} {\mathop {\tilde Q}\nolimits_{F,a} ^{(-2a + N + m)(\left| {\mathop \mu \nolimits_{a + 1} } \right| +  \cdots  + \left| {\mathop \mu \nolimits_N } \right|)} } \nonumber\\
&= \left( {\prod\limits_{a = 1}^{\frac{{N - 1}}{2}} {\prod\limits_{b = a}^{\frac{{N - 1}}{2}} {\mathop {\tilde Q}\nolimits_{F,b} ^{\left| {\mathop \mu \nolimits_a } \right|} } } \prod\limits_{a = \frac{{N + 1}}{2} + 1}^{N - 1} {\prod\limits_{b = \frac{{N + 1}}{2}}^{a - 1} {\mathop {\tilde Q}\nolimits_{F,b} ^{ - \left| {\mathop \mu \nolimits_a } \right|} } } } \right)^m \left( {\prod\limits_{a = \frac{{N + 1}}{2}}^{\frac{{N + m-1}}{2}} {\mathop {\tilde Q}\nolimits_{F,a} ^{(N + m - 2a)} } } \right)^{(\left| {\mathop \mu \nolimits_1 } \right| +  \cdots  + \left| {\mathop \mu \nolimits_N } \right|)} \nonumber\\
 &=
\mathop {\left({{\mathop {\left( {\mathop e\nolimits_{\frac{{N + 1}}{2}} } \right)}\nolimits^{-m} }{\prod\limits_{a = \frac{{N + 1}}{2}}^{\frac{{N + m - 1}}{2}} {\mathop {\tilde Q}\nolimits_{F,a} ^{(N + m - 2a)} } }} \right)}\nolimits^{\left| {\vec \mu } \right|} \prod\limits_{a = 1}^{N - 1} {\mathop e\nolimits_a ^{m\left| {\mathop \mu \nolimits_a } \right|} } 
\end{align}
%(\ref {id:3-4})(\ref {id:3-5})
(\ref {formula01})(\ref {formula1})(\ref {formula2}) provide the identity (\ref {id:3-3}) immediately.

Finally, (\ref{nek:rewritten})
(\ref {id:3-3}) imply the following equality
\begin{align}
\label{equal}
\mathop Z\nolimits_{inst.}^{A - model,SU(N)} \left( {\mathop Q\nolimits_B ,\mathop Q\nolimits_{F,a} } \right) = \mathop Z\nolimits_{inst.}^{Nek,SU(N)} \left( {\hat Q,\mathop Q\nolimits_{ab} } \right)
\end{align}
where $\hat Q = (-1)^{N}C(\mathop e\nolimits_a ,t,q,m)\mathop Q\nolimits_B$. 

Note that if we use the framing factor without our modification, we get the additional factor 
\begin{align}
\prod\limits_{a = 1}^N {\left( {\frac{t}{q}} \right)^{ - \frac{{\mathop n\nolimits_a \left| {\mathop \mu \nolimits_a } \right|}}{2}} } 
\end{align}
in the summation of the refined patririon function. They cannot be absorbed into the ${\hat Q}$ and break the equivalence (\ref{equal}). Hence we need the modification (\ref{modification}).

%%%%%%%%%%%%%%%%%%%%%%%%%%%%%%%%%%%%%%%%%%%%%%%%%%%%%%%%%%%%%%%%%%
\subsection{Adding Matters and Strip Geometries}
\label{ssec:3.2}
\begin{figure}[h]
\begin{center}
%WinTpicVersion3.08
\unitlength 0.1in
\begin{picture}( 38.8900, 11.7100)( -6.8200,-14.5100)
% LINE 2 2 3 0
% 10 452 486 452 1240 452 1240 2665 1240 452 486 2660 486 1220 1240 1220 491 1220 491 452 1240
% 
\special{pn 8}%
\special{pa 452 486}%
\special{pa 452 1240}%
\special{dt 0.050}%
\special{pa 452 1240}%
\special{pa 2666 1240}%
\special{dt 0.050}%
\special{pa 452 486}%
\special{pa 2660 486}%
\special{dt 0.050}%
\special{pa 1220 1240}%
\special{pa 1220 492}%
\special{dt 0.050}%
\special{pa 1220 492}%
\special{pa 452 1240}%
\special{dt 0.050}%
% LINE 2 2 3 0
% 6 2650 486 3145 486 2761 1240 1225 486 2655 1240 3140 1240
% 
\special{pn 8}%
\special{pa 2650 486}%
\special{pa 3146 486}%
\special{dt 0.050}%
\special{pa 2762 1240}%
\special{pa 1226 486}%
\special{dt 0.050}%
\special{pa 2656 1240}%
\special{pa 3140 1240}%
\special{dt 0.050}%
% LINE 2 2 3 0
% 4 1988 1240 1220 491 1993 486 2761 1240
% 
\special{pn 8}%
\special{pa 1988 1240}%
\special{pa 1220 492}%
\special{dt 0.050}%
\special{pa 1994 486}%
\special{pa 2762 1240}%
\special{dt 0.050}%
% LINE 1 0 3 0
% 12 630 904 798 1106 798 1106 1719 1106 1719 1106 1921 904 1921 904 1983 770 1983 770 2209 544 2209 544 2209 285
% 
\special{pn 13}%
\special{pa 630 904}%
\special{pa 798 1106}%
\special{fp}%
\special{pa 798 1106}%
\special{pa 1720 1106}%
\special{fp}%
\special{pa 1720 1106}%
\special{pa 1922 904}%
\special{fp}%
\special{pa 1922 904}%
\special{pa 1984 770}%
\special{fp}%
\special{pa 1984 770}%
\special{pa 2210 544}%
\special{fp}%
\special{pa 2210 544}%
\special{pa 2210 286}%
\special{fp}%
% LINE 1 0 3 0
% 12 620 899 620 280 620 899 260 899 798 1101 798 1442 1719 1110 1719 1446 1926 904 1926 1451 1978 755 1978 285
% 
\special{pn 13}%
\special{pa 620 900}%
\special{pa 620 280}%
\special{fp}%
\special{pa 620 900}%
\special{pa 260 900}%
\special{fp}%
\special{pa 798 1102}%
\special{pa 798 1442}%
\special{fp}%
\special{pa 1720 1110}%
\special{pa 1720 1446}%
\special{fp}%
\special{pa 1926 904}%
\special{pa 1926 1452}%
\special{fp}%
\special{pa 1978 756}%
\special{pa 1978 286}%
\special{fp}%
% LINE 2 2 3 0
% 2 2761 1245 3145 866
% 
\special{pn 8}%
\special{pa 2762 1246}%
\special{pa 3146 866}%
\special{dt 0.050}%
% LINE 1 0 3 0
% 6 2209 539 3054 539 3054 539 3207 693 3054 539 3054 290
% 
\special{pn 13}%
\special{pa 2210 540}%
\special{pa 3054 540}%
\special{fp}%
\special{pa 3054 540}%
\special{pa 3208 694}%
\special{fp}%
\special{pa 3054 540}%
\special{pa 3054 290}%
\special{fp}%
% LINE 2 2 3 0
% 2 2766 1240 2766 491
% 
\special{pn 8}%
\special{pa 2766 1240}%
\special{pa 2766 492}%
\special{dt 0.050}%
% STR 2 0 3 0
% 3 2910 1070 2910 1150 3 0
% $(-1,-1)$ curve
\put(29.1000,-11.5000){\makebox(0,0)[rb]{$(-1,-1)$ curve}}%
% STR 2 0 3 0
% 3 1190 910 1190 990 5 0
% $(-2,0)$ curve
\put(11.9000,-9.9000){\makebox(0,0){$(-2,0)$ curve}}%
% STR 2 0 3 0
% 3 668 1024 668 1104 3 0
% $(-1,-1)$ curve
\put(6.6800,-11.0400){\makebox(0,0)[rb]{$(-1,-1)$ curve}}%
% STR 2 0 3 0
% 3 3020 860 3020 940 3 0
% $(-1,-1)$ curve
\put(30.2000,-9.4000){\makebox(0,0)[rb]{$(-1,-1)$ curve}}%
\end{picture}%
\end{center}
\caption{A toric diagram of a strip geometry which is obtained from triangulation of a strip toric data}
\label{strip}
\end{figure}
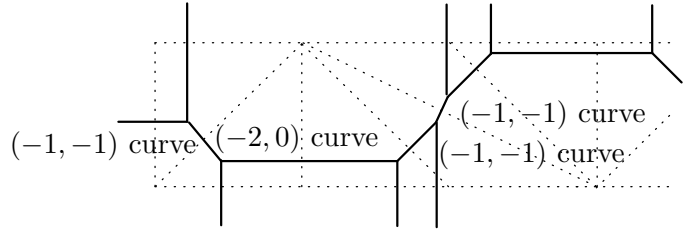

%%%%%%%%%%%%%%%%%%%%%%%%%%%%%%%%%%%%%%%%%%%%%%%%%%%%%%%%%%%%%%%%%%
By blowing up the $SU(N)$ geometries, we can add matters to the Nekrasov's instanton calculation via the geometric engineering. The K\"{a}hler parameters of the blown up $\mathop {\mathbb P}^{1}$'s give rise to the mass parameters of the matters. These geometries is obtained by gluing strip geometries. A strip geometry is a toric Calabi-Yau that contains a chain of $\mathop {\mathbb P}^{1}$'s. Each $\mathop {\mathbb P}^{1}$ locally forms a $(-1,-1)$ curve $O( - 1) \oplus O( - 1) \to \mathop {\mathbb P}^{1}
$ or $(-2,0)$ curve $O( - 2) \oplus O( 0) \to \mathop {\mathbb P}^{1}
$ as Fig.\ref {strip}.
%%%%%%%%%%%%%%%%%%%%%%%%%%%%%%%%%%%%%%%%%%%%%%%%%%%%%%%%%%%%%%%%%%%%%%%%%%%%%%%%%%%%%%%%%%%%%%%%%%%%%%%%%%%%%%%%%%%%%%%%%%%%%%%%%%%%%%%%%%%%%%%%%%%%%%%%%%%%%%%
Following \cite{Iqbal:2004ne}, we take the chain of $(-1,-1)$ curves Fig.\ref{2N} for example. Gluing these strip geometries, we get the toric Calabi-Yau that engineers $\cal N$$=2$ $SU(N)$ gauge theory with $\mathop N\nolimits_f  = 2N $ \cite{Iqbal:2004ne}. The refined vertex on the strip geometry Fig.\ref{2N}(a) yields
\begin{align}
\mathop K\nolimits_{\mathop \beta \nolimits_1 \mathop \beta \nolimits_2  \cdots }^{\mathop \alpha \nolimits_1 \mathop \alpha \nolimits_{2 \cdots } }  &= \sum\limits_{\left\{ {\mathop \mu \nolimits_a } \right\},\left\{ {\mathop \nu \nolimits_a } \right\}} {\left( { - \mathop Q\nolimits_{M,1} } \right)^{\left| {\mathop \mu \nolimits_1 } \right|}  \left( { - \mathop Q\nolimits_{F,2} } \right)^{\left| {\mathop \nu \nolimits_2 } \right|} \left( { - \mathop Q\nolimits_{M,2} } \right)^{\left| {\mathop \mu \nolimits_2 } \right|}  \cdots } \cr
& \times \mathop C\nolimits_{\mathop {\mathop \nu \nolimits_1^t }
 \mathop \mu \nolimits_1 \mathop \alpha \nolimits_1 } \mathop C\nolimits_{\mathop \nu \nolimits_2 \mathop \mu \nolimits_1 ^t \mathop \beta \nolimits_1 } \mathop C\nolimits_{\mathop \nu \nolimits_2 ^t \mathop \mu \nolimits_2 \mathop \alpha \nolimits_2 } \mathop C\nolimits_{\mathop \nu \nolimits_3 \mathop \mu \nolimits_2 ^t \mathop \beta \nolimits_2 }  \times  \cdots \nonumber
 \end{align}
 \begin{align}
 &= \prod\limits_a {\left[ {\mathop q\nolimits^{\frac{{\left\| {\mathop \alpha \nolimits_a } \right\|^2 }}{2}} \mathop t\nolimits^{\frac{{\left\| {\mathop \beta \nolimits_a } \right\|^2 }}{2}} \mathop {\tilde Z}\nolimits_{\mathop \alpha \nolimits_a } (t,q)\mathop {\tilde Z}\nolimits_{\mathop \beta \nolimits_a } (q,t)} \right]} 
\cr
& \times \sum\limits_{
   {\left\{ {\mathop \mu \nolimits_a } \right\},\left\{ {\mathop \nu \nolimits_a } \right\}}  \atop
   {\left\{ {\mathop \rho \nolimits_a } \right\},\left\{ {\mathop \sigma \nolimits_a } \right\}}  
} {\prod\limits_a {\left( { - \mathop Q\nolimits_{M,a} } \right)^{\left| {\mathop \mu \nolimits_a } \right|} \left( { - \mathop Q\nolimits_{F,a} } \right)^{\left| {\mathop \nu \nolimits_a } \right|} \mathop {\left( {\frac{q}{t}} \right)}\nolimits^{\frac{{\left\| {\mathop \mu \nolimits_a } \right\|^2 }}{2}} \mathop t\nolimits^{\frac{{\mathop \kappa \nolimits_{\mathop \mu \nolimits_a } }}{2}} \mathop {\left( {\frac{t}{q}} \right)}\nolimits^{\frac{{|| {\mathop \mu \nolimits_a^t } ||^2 }}{2}} \mathop q\nolimits^{ - \frac{{\mathop \kappa \nolimits_{\mathop \mu \nolimits_a } }}{2}} } } \nonumber\\
& \times \mathop {\left( {\frac{q}{t}} \right)}\nolimits^{\frac{{\left| {\mathop \rho \nolimits_a } \right| + \left| {\mathop \nu \nolimits_a } \right| - \left| {\mathop \mu \nolimits_a } \right|}}{2}} \mathop {\left( {\frac{t}{q}} \right)}\nolimits^{\frac{{\left| {\mathop \sigma \nolimits_a } \right| + \left| {\mathop \nu \nolimits_{a + 1} } \right| - \left| {\mathop \mu \nolimits_a } \right|}}{2}} \nonumber\\
& \times \mathop s\nolimits_{\mathop \nu \nolimits_a /\mathop \rho \nolimits_a } \left( {\mathop t\nolimits^{ - \rho } \mathop q\nolimits^{ - \mathop \alpha \nolimits_a } } \right)\mathop s\nolimits_{\mathop \mu \nolimits_a /\mathop \rho \nolimits_a } \left( {\mathop t\nolimits^{ - \mathop \alpha \nolimits_a^t } \mathop q\nolimits^{ - \rho } } \right)\mathop s\nolimits_{\mathop \nu \nolimits_{a + 1}^t /\mathop \sigma \nolimits_a } \left( {\mathop q\nolimits^{ - \rho } \mathop t\nolimits^{ - \mathop \beta \nolimits_a } } \right)\mathop s\nolimits_{\mathop \mu \nolimits_a^t /\mathop \sigma \nolimits_a } \left( {\mathop q\nolimits^{ - \mathop \beta \nolimits_a^t } \mathop t\nolimits^{ - \rho } } \right)
\end{align}
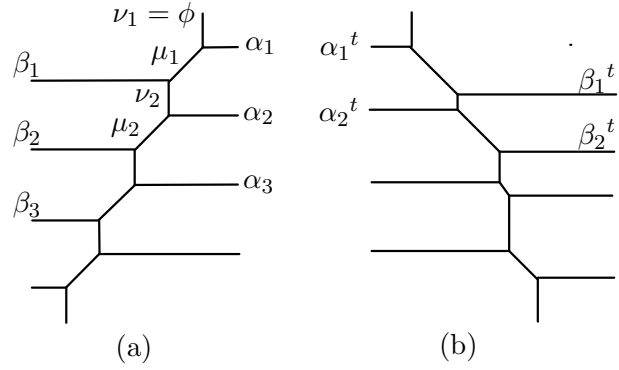
\begin{figure}[h]
\begin{center}
%WinTpicVersion3.08
\unitlength 0.1in
\begin{picture}( 34.8700, 17.5400)(  1.2300,-18.1900)
% STR 2 0 3 0
% 3 1740 620 1740 680 5 0
% $\alpha_2$
\put(17.4000,-6.8000){\makebox(0,0){$\alpha_2$}}%
% LINE 1 0 3 0
% 34 1446 146 1446 323 1446 323 1629 322 1449 320 1267 505 1267 505 1267 679 1267 679 1630 679 1270 679 1090 861 1090 861 1090 1044 1090 1044 1630 1042 1093 1041 904 1226 904 1226 905 1406 905 1406 1637 1405 905 1403 731 1580 731 1580 731 1766 728 1581 551 1581 901 1229 553 1230 1090 858 550 858 1267 496 550 498
% 
\special{pn 13}%
\special{pa 1446 146}%
\special{pa 1446 324}%
\special{fp}%
\special{pa 1446 324}%
\special{pa 1630 322}%
\special{fp}%
\special{pa 1450 320}%
\special{pa 1268 506}%
\special{fp}%
\special{pa 1268 506}%
\special{pa 1268 680}%
\special{fp}%
\special{pa 1268 680}%
\special{pa 1630 680}%
\special{fp}%
\special{pa 1270 680}%
\special{pa 1090 862}%
\special{fp}%
\special{pa 1090 862}%
\special{pa 1090 1044}%
\special{fp}%
\special{pa 1090 1044}%
\special{pa 1630 1042}%
\special{fp}%
\special{pa 1094 1042}%
\special{pa 904 1226}%
\special{fp}%
\special{pa 904 1226}%
\special{pa 906 1406}%
\special{fp}%
\special{pa 906 1406}%
\special{pa 1638 1406}%
\special{fp}%
\special{pa 906 1404}%
\special{pa 732 1580}%
\special{fp}%
\special{pa 732 1580}%
\special{pa 732 1766}%
\special{fp}%
\special{pa 728 1582}%
\special{pa 552 1582}%
\special{fp}%
\special{pa 902 1230}%
\special{pa 554 1230}%
\special{fp}%
\special{pa 1090 858}%
\special{pa 550 858}%
\special{fp}%
\special{pa 1268 496}%
\special{pa 550 498}%
\special{fp}%
% STR 2 0 3 0
% 3 1086 1844 1086 1904 5 0
% (a)
\put(10.8600,-19.0400){\makebox(0,0){(a)}}%
% STR 2 0 3 0
% 3 2784 1832 2784 1892 5 0
% (b)
\put(27.8400,-18.9200){\makebox(0,0){(b)}}%
% STR 2 0 3 0
% 3 1750 255 1750 315 5 0
% $\alpha_1$
\put(17.5000,-3.1500){\makebox(0,0){$\alpha_1$}}%
% STR 2 0 3 0
% 3 1740 980 1740 1040 5 0
% $\alpha_3$
\put(17.4000,-10.4000){\makebox(0,0){$\alpha_3$}}%
% STR 2 0 3 0
% 3 528 356 528 416 5 0
% $\beta_1$
\put(5.2800,-4.1600){\makebox(0,0){$\beta_1$}}%
% STR 2 0 3 0
% 3 528 716 528 776 5 0
% $\beta_2$
\put(5.2800,-7.7600){\makebox(0,0){$\beta_2$}}%
% STR 2 0 3 0
% 3 528 1088 528 1148 5 0
% $\beta_3$
\put(5.2800,-11.4800){\makebox(0,0){$\beta_3$}}%
% STR 2 0 3 0
% 3 1190 90 1190 150 5 0
% ${\nu_1} = \phi$
\put(11.9000,-1.5000){\makebox(0,0){${\nu_1} = \phi$}}%
% STR 2 0 3 0
% 3 1158 530 1158 590 5 0
% $\nu_2$
\put(11.5800,-5.9000){\makebox(0,0){$\nu_2$}}%
% STR 2 0 3 0
% 3 1254 308 1254 368 5 0
% $\mu_1$
\put(12.5400,-3.6800){\makebox(0,0){$\mu_1$}}%
% STR 2 0 3 0
% 3 1044 686 1044 746 5 0
% $\mu_2$
\put(10.4400,-7.4600){\makebox(0,0){$\mu_2$}}%
% STR 2 0 3 0
% 3 2164 255 2164 315 5 0
% ${\alpha_1}^t$
\put(21.6400,-3.1500){\makebox(0,0){${\alpha_1}^t$}}%
% STR 2 0 3 0
% 3 3510 430 3510 490 5 0
% ${\beta_1}^t$
\put(35.1000,-4.9000){\makebox(0,0){${\beta_1}^t$}}%
% STR 2 0 3 0
% 3 2160 600 2160 660 5 0
% ${\alpha_2}^t$
\put(21.6000,-6.6000){\makebox(0,0){${\alpha_2}^t$}}%
% STR 2 0 3 0
% 3 3510 720 3510 780 5 0
% ${\beta_2}^t$
\put(35.1000,-7.8000){\makebox(0,0){${\beta_2}^t$}}%
% LINE 1 0 3 0
% 2 3370 310 3370 310
% 
\special{pn 13}%
\special{pa 3370 310}%
\special{pa 3370 310}%
\special{fp}%
% LINE 1 0 3 0
% 24 2330 320 2530 320 2530 320 2780 570 2780 570 2780 650 2780 650 2320 650 2780 650 3000 870 3000 870 3000 1030 3000 1030 2330 1030 3000 1030 3050 1100 3050 1100 3050 1390 3050 1390 2330 1390 3050 1390 3200 1540 3200 1540 3200 1760
% 
\special{pn 13}%
\special{pa 2330 320}%
\special{pa 2530 320}%
\special{fp}%
\special{pa 2530 320}%
\special{pa 2780 570}%
\special{fp}%
\special{pa 2780 570}%
\special{pa 2780 650}%
\special{fp}%
\special{pa 2780 650}%
\special{pa 2320 650}%
\special{fp}%
\special{pa 2780 650}%
\special{pa 3000 870}%
\special{fp}%
\special{pa 3000 870}%
\special{pa 3000 1030}%
\special{fp}%
\special{pa 3000 1030}%
\special{pa 2330 1030}%
\special{fp}%
\special{pa 3000 1030}%
\special{pa 3050 1100}%
\special{fp}%
\special{pa 3050 1100}%
\special{pa 3050 1390}%
\special{fp}%
\special{pa 3050 1390}%
\special{pa 2330 1390}%
\special{fp}%
\special{pa 3050 1390}%
\special{pa 3200 1540}%
\special{fp}%
\special{pa 3200 1540}%
\special{pa 3200 1760}%
\special{fp}%
% LINE 1 0 3 0
% 2 2540 320 2540 140
% 
\special{pn 13}%
\special{pa 2540 320}%
\special{pa 2540 140}%
\special{fp}%
% LINE 1 0 3 0
% 2 2780 570 3610 570
% 
\special{pn 13}%
\special{pa 2780 570}%
\special{pa 3610 570}%
\special{fp}%
% LINE 1 0 3 0
% 6 3000 870 3600 870 3050 1100 3590 1100 3200 1530 3600 1530
% 
\special{pn 13}%
\special{pa 3000 870}%
\special{pa 3600 870}%
\special{fp}%
\special{pa 3050 1100}%
\special{pa 3590 1100}%
\special{fp}%
\special{pa 3200 1530}%
\special{pa 3600 1530}%
\special{fp}%
\end{picture}%
\end{center}
\caption{The building blocks of the toric Calabi-Yau that engineers $SU(N)$ gauge theory with $\mathop N\nolimits_f  = 2N $}
\label{2N}
\end{figure}
where $\mathop \nu \nolimits_1  = \mathop \nu \nolimits_{N + 1}  = \phi $. They involve the following sum
\begin{align}
&\sum\limits_{
   {\left\{ {\mathop \mu \nolimits_a } \right\},\left\{ {\mathop \nu \nolimits_a } \right\}}  \atop
   {\left\{ {\mathop \rho \nolimits_a } \right\},\left\{ {\mathop \sigma \nolimits_a } \right\}}  
} {\prod\limits_a {\left( { - {\mathop Q\nolimits_{M,a} }} \right)^{\left| {\mathop \mu \nolimits_a } \right|} \left( { - \mathop Q\nolimits_{F,a} } \right)^{\left| {\mathop \nu \nolimits_a } \right|} } } \cr
& \times \mathop s\nolimits_{\mathop \nu \nolimits_a /\mathop \rho \nolimits_a } \left( {\mathop t\nolimits^{ - \rho + \frac{1}{2}} \mathop q\nolimits^{ - \mathop \alpha \nolimits_a  } } \right)\mathop s\nolimits_{\mathop \mu \nolimits_a /\mathop \rho \nolimits_a } \left( {\mathop t\nolimits^{ - \mathop \alpha \nolimits_a^t  } \mathop q\nolimits^{ - \rho - \frac{1}{2}} } \right)\mathop s\nolimits_{\mathop \nu \nolimits_{a + 1}^t /\mathop \sigma \nolimits_a } \left( {\mathop q\nolimits^{ - \rho } \mathop t\nolimits^{ - \mathop \beta \nolimits_a  - \frac{1}{2}} } \right)
\mathop s\nolimits_{\mathop \mu \nolimits_a^t /\mathop \sigma \nolimits_a } \left( {\mathop q\nolimits^{ - \mathop \beta \nolimits_a^t  + \frac{1}{2}} \mathop t\nolimits^{ - \rho } } \right) \nonumber
\end{align}
Using the method of Iqbal-KashaniPoor\cite{Iqbal:2004ne}, we can take the summation. The only difference from the result of \cite{Iqbal:2004ne} is the arguments of Schur functions. Bewaring the difference, we get the sum as follows
\begin{align}
&\prod\limits_{1 \le a \le b \le N} {\mathop {\left[ {\mathop t\nolimits^{ - \mathop \alpha \nolimits_a^t } \mathop q\nolimits^{ - \rho  - \frac{1}{2}} ,{ - \mathop Q\nolimits_{\mathop \alpha \nolimits_a \mathop \beta \nolimits_b } } \mathop q\nolimits^{ - \mathop \beta \nolimits_b^t  + \frac{1}{2}} \mathop t\nolimits^{ - \rho } } \right]}} \prod\limits_{1 \le a < b \le N} {\mathop {\left[ {\mathop q\nolimits^{ - \rho } \mathop t\nolimits^{ - \mathop \beta \nolimits_a  - \frac{1}{2}} ,{ - \mathop Q\nolimits_{\mathop \beta \nolimits_a \mathop \alpha \nolimits_b } }\mathop t\nolimits^{ - \rho  + \frac{1}{2}} \mathop q\nolimits^{ - \mathop \alpha \nolimits_b } } \right]} } \nonumber\\
& \times \prod\limits_{1 \le a < b \le N} {\mathop {\left\{  {\mathop t\nolimits^{ - \mathop \alpha \nolimits_a^t } \mathop q\nolimits^{ - \rho  - \frac{1}{2}} ,{\mathop Q\nolimits_{\mathop \alpha \nolimits_a \mathop \alpha \nolimits_b }}\mathop t\nolimits^{ - \rho  + \frac{1}{2}} \mathop q\nolimits^{ - \mathop \alpha \nolimits_b } } \right\} } \mathop {\left\{ {\mathop q\nolimits^{ - \rho } \mathop t\nolimits^{ - \mathop \beta \nolimits_a  - \frac{1}{2}} ,{\mathop Q\nolimits_{\mathop \beta \nolimits_a \mathop \beta \nolimits_b } }\mathop t\nolimits^{ - \rho } \mathop q\nolimits^{ - \mathop \beta \nolimits_b^t  + \frac{1}{2}} } \right\}} } 
\end{align}
We provide the direct proof in appendix \ref{exam}. Here K\"{a}hler parameters are given by
\begin{eqnarray*}
   \mathop Q\nolimits_{\mathop \alpha \nolimits_a \mathop \beta \nolimits_b }  &=& \mathop Q\nolimits_{M,a} \mathop Q\nolimits_{F,a+1}  \cdots \mathop Q\nolimits_{M,b - 1} \mathop Q\nolimits_{F,b } \mathop Q\nolimits_{M,b}  = \mathop Q\nolimits_{a,b} \mathop Q\nolimits_{M,b}   \\
   \mathop Q\nolimits_{\mathop \beta \nolimits_a \mathop \alpha \nolimits_b }  &=& \mathop Q\nolimits_{F,a+1}  \cdots \mathop Q\nolimits_{M,b-1 } \mathop Q\nolimits_{F,b}  = \mathop Q\nolimits_{M,a}^{ - 1} \mathop Q\nolimits_{a,b}   \\
   \mathop Q\nolimits_{\mathop \alpha \nolimits_a \mathop \alpha \nolimits_b }  &=& \mathop Q\nolimits_{M,a} \mathop Q\nolimits_{F,a+1}  \cdots \mathop Q\nolimits_{M,b - 1} \mathop Q\nolimits_{F,b }  = \mathop Q\nolimits_{a,b}   \\
   \mathop Q\nolimits_{\mathop \beta \nolimits_a \mathop \beta \nolimits_b }  &=& \mathop Q\nolimits_{F,a+1}  \cdots \mathop Q\nolimits_{M,b - 1} \mathop Q\nolimits_{F,b } \mathop Q\nolimits_{M,b}  = \mathop Q\nolimits_{M,a}^{ - 1} \mathop Q\nolimits_{a,b} \mathop Q\nolimits_{M,b}   \\
\end{eqnarray*}
and we introduce 
\begin{align}
\left[ {x,y} \right] \equiv \prod\limits_{i,j = 1}^\infty  {\left( {1 + \mathop x\nolimits_i \mathop y\nolimits_j } \right)} ,  \left\{ {x,y} \right\} \equiv \prod\limits_{i,j = 1}^\infty  {\mathop {\left( {1 - \mathop x\nolimits_i \mathop y\nolimits_j } \right)}\nolimits^{ - 1} } 
\end{align}
Then we obtain the following expression
\begin{align}
\mathop K\nolimits_{\mathop \beta \nolimits_1 \mathop \beta \nolimits_2  \cdots }^{\mathop \alpha \nolimits_1 \mathop \alpha \nolimits_{2 \cdots } } 
&= \prod\limits_a {\left[ {\mathop q\nolimits^{\frac{{\left\| {\mathop \alpha \nolimits_a } \right\|^2 }}{2}} \mathop t\nolimits^{\frac{{\left\| {\mathop \beta \nolimits_a } \right\|^2 }}{2}} \mathop {\tilde Z}\nolimits_{\mathop \alpha \nolimits_a } (t,q)\mathop {\tilde Z}\nolimits_{\mathop \beta \nolimits_a } (q,t)} \right]} \cr
&\times \prod\limits_{i,j = 1}^\infty  {\prod\limits_{1 \le a \le b \le N} {\left( {1 - \mathop Q\nolimits_{\mathop \alpha \nolimits_a \mathop \beta \nolimits_b } \mathop t\nolimits^{ - \mathop \alpha \nolimits_{a,i} ^t  + j - \frac{1}{2}} \mathop q\nolimits^{ - \mathop \beta \nolimits_{b,j} ^t  + i - \frac{1}{2}} } \right)} } \prod\limits_{1 \le a < b \le N} {\left( {1 - \mathop Q\nolimits_{\mathop \beta \nolimits_a \mathop \alpha \nolimits_b } \mathop t\nolimits^{ - \mathop \beta \nolimits_{a,i}  + j - \frac{1}{2}} \mathop q\nolimits^{ - \mathop \alpha \nolimits_{b,j}  + i - \frac{1}{2}} } \right)} \cr
&\times \prod\limits_{1 \le a < b \le N} {\left( {1 - \mathop Q\nolimits_{\mathop \alpha \nolimits_a \mathop \alpha \nolimits_b } \mathop t\nolimits^{ - \mathop \alpha \nolimits_{a,i} ^t  + j} \mathop q\nolimits^{ - \mathop \alpha \nolimits_{b,j}  + i - 1} } \right)^{ - 1} \left( {1 - \mathop Q\nolimits_{\mathop \beta \nolimits_a \mathop \beta \nolimits_b } \mathop t\nolimits^{ - \mathop \beta \nolimits_{a,i}  + j - 1} \mathop q\nolimits^{ - \mathop \beta \nolimits_{b,j} ^t  + i} } \right)^{ - 1} } 
\end{align}
The amplitude for the pair of this strip geomerty Fig.\ref{2N}(b) is given by
\begin{align}
\mathop {\tilde K}\nolimits_{\mathop \alpha \nolimits_1 ^t \mathop \alpha \nolimits_2 ^t  \cdots }^{\mathop \beta \nolimits_1 ^t \mathop \beta \nolimits_2 ^t  \cdots } 
&= \prod\limits_a {\left[ {\mathop t\nolimits^{\frac{{|| {{\alpha}^t_a}||^2 }}{2}} \mathop q\nolimits^{\frac{{|| {\mathop {\beta}^t \nolimits_a }||^2 }}{2}} \mathop {\tilde Z}\nolimits_{\mathop {\alpha}^t \nolimits_a } (q,t)\mathop {\tilde Z}\nolimits_{\mathop {\beta}^t \nolimits_a } (t,q)} \right]} \cr
&\times \prod\limits_{i,j = 1}^\infty  {\prod\limits_{1 \le a \le b \le N} {\left( {1 - \mathop Q'\nolimits_{\mathop \alpha \nolimits_a \mathop \beta \nolimits_b } \mathop t\nolimits^{ - \mathop \alpha \nolimits_{a,i} ^t  + j - \frac{1}{2}} \mathop q\nolimits^{ - \mathop \beta \nolimits_{b,j} ^t  + i - \frac{1}{2}} } \right)} } \prod\limits_{1 \le a < b \le N} {\left( {1 - \mathop Q'\nolimits_{\mathop \beta \nolimits_a \mathop \alpha \nolimits_b } \mathop t\nolimits^{ - \mathop \beta \nolimits_{a,i}  + j - \frac{1}{2}} \mathop q\nolimits^{ - \mathop \alpha \nolimits_{b,j}  + i - \frac{1}{2}} } \right)} \cr
&\times \prod\limits_{1 \le a < b \le N} {\left( {1 - \mathop Q'\nolimits_{\mathop \alpha \nolimits_a \mathop \alpha \nolimits_b } \mathop t\nolimits^{ - \mathop \alpha \nolimits_{a,i} ^t  + j-1} \mathop q\nolimits^{ - \mathop \alpha \nolimits_{b,j}  + i } } \right)^{ - 1} \left( {1 - \mathop Q'\nolimits_{\mathop \beta \nolimits_a \mathop \beta \nolimits_b } \mathop t\nolimits^{ - \mathop \beta \nolimits_{a,i}  + j } \mathop q\nolimits^{ - \mathop \beta \nolimits_{b,j} ^t  + i-1} } \right)^{ - 1} } 
\end{align}
Gluing them, we get the Nekrasov's partition functioin for $\cal N$$=2$ $SU(N)$ gauge theory with $\mathop N\nolimits_f  = 2N $ 
\begin{eqnarray*}
Z = \sum\limits_{\mathop \alpha \nolimits_1 \mathop \alpha \nolimits_2  \cdots } {\prod\limits_{a = 1}^N {\left( {\mathop f\nolimits_{\mathop \alpha \nolimits_a } (t,q)\mathop {\mathop Q\nolimits_B }\nolimits^{\left| {\mathop \alpha \nolimits_a } \right|} } \right)} \mathop {K}\nolimits_{\phi \phi  \cdots  \cdots }^{\mathop \alpha \nolimits_1 \mathop \alpha \nolimits_2  \cdots }(\mathop Q\nolimits_{ab} ,\mathop Q\nolimits_{M,a} ) \mathop {\tilde K}\nolimits_{\mathop \alpha \nolimits_1 ^t \mathop \alpha \nolimits_2 ^t  \cdots }^{\phi \phi  \cdots } (\mathop Q\nolimits_{ab} ,\mathop Q'\nolimits_{M,a} )} 
\end{eqnarray*}
It is not so hard to generarize the above caluculation of the refined vertex for another strip geometries which contain $(-1,-1)$ curves and $(-2,0)$ curves. Then we can engineer the Nekrasov's partition functioins for various $\cal N$$=2$ $SU(N)$ quiver gauge theories with matters by gluing these amplituses.

%%%%%%%%%%%%%%%%%%%%%%%%%%%%%%%%%%%%%%%%%%%%%%%%
\section{Conclusion}
%%%%%%%%%%%%%%%%%%%%%%%%%%%%%%%%%%%%%%%%%%%%%%%%

In this paper, we have applied refined topological vertex for $SU(N)$ geometries and reproduced the K-theoretic version of the Nekrasov's partition functions.
From this results we can adopt refined topological vertex as a 2-parameter extension of topological A-model.
We have also discussed a refined vertex on a strip geometry. Many of the nice properties obtained in \cite{Iqbal:2004ne} are maintained in the case of refined vertex. The important point is that refined vertex on strip reduces to a summation of Schur functions which is essentially discussed in \cite{Iqbal:2004ne}. Hence Schur functions of the partition functions can be summed up as in the case of the topological vertex on strips.

%%%%%%%%%%%%%%%%%%%%%%%%%%%%%%%%%%%%%%%%%%%%%%%%%%%%%%%%%
\section*{Acknowledgements}

We would like to thank Tohru Eguchi, Yosuke Imamura, Hiroaki Kanno and Yuji Tachikawa for valuable discussions and helpful comments.
%%%%%%%%%%%%%%%%%%%%%%%%%%%%%%%%%%%%%%%%%%%%%%%%%%%%%%%%%

%%%%%%%%%%%%%%%%%%%%%%%%%%%%%%%%%%%%%%%%%%%%%%%%%%%%%%%%%%
\appendix
%%%%%%%%%%%%%%%%%%%%%%%%%%%%%%%%%%%%%%%%%%%%%%%%%%%%%%%%%%
\section{Young diagrams and Schur functions }
\label{appendix:a}
%%%%%%%%%%%%%%%%%%%%%%%%%%%%%%%%%%%%%%%%%%%%%%%%%%%%%%%%%%
{\bf Young diagrams}

The Young diagrams is defined as a sequence of decreasing non-negative integers 
\begin{eqnarray}
\mu  = \left\{ {\mathop \mu \nolimits_i  \in \mathop {\mathbb Z}\nolimits_{ \ge 0} |\mathop \mu \nolimits_1  \ge \mathop \mu \nolimits_2  \ge  \cdots } \right\}\end{eqnarray}
The transpose of $\mu $ is defined as follows
\begin{eqnarray}
\mathop \mu \nolimits^t  = \left\{ {\mathop \mu \nolimits_j^t  \in \mathop Z\nolimits_{ \ge 0} |\mathop \mu \nolimits_j^t  = \# \left\{ {i|\mathop \mu \nolimits_i  \ge j} \right\}} \right\}
\end{eqnarray}
The size and the norm of the partition is denoted as
\begin{eqnarray}
\left| \mu  \right| = \sum\limits_{i = 1}^{d(\mu )} {\mathop \mu \nolimits_i } ,\quad   \left\| \mu  \right\|^2  = \sum\limits_{i = 1}^{d(\mu )} {\mathop \mu \nolimits_i^2 } 
\end{eqnarray}
For $(i,j)  \in \mu $, we define the following quantities,
\begin{eqnarray*}
   {\mathop a\nolimits_\mu  (i,j) = \mathop \mu \nolimits_i  - j} , & {\mathop l\nolimits_\mu  (i,j) = \mathop \mu \nolimits_j^t  - i}  
\end{eqnarray*}
\begin{eqnarray*}
   {\mathop {a'}\nolimits_\mu  (i,j) = j - 1} , & {\mathop {l'}\nolimits_\mu  (i,j) = i - 1}  
\end{eqnarray*}
We introduce the hook length of the Young diagram
\begin{eqnarray*}
{\mathop h\nolimits_\mu  (i,j) = \mathop \mu \nolimits_i  - j + \mathop \mu \nolimits_j^t  - i + 1}
\end{eqnarray*}
It is also useful to define the following quantities
\begin{eqnarray*}
\begin{array}{*{20}c}
   {n(\mu ) = \sum\limits_{i = 1}^{d(\mu )} {(i - 1)\mathop \mu \nolimits_i } } & {\mathop \kappa \nolimits_\mu   = \sum\limits_{(i,j) \in \mu } {(j - i)} }  \\
\end{array}
\end{eqnarray*}
It is easy to show that they satisfy the following identities
\begin{eqnarray}
\label{A1}
{n(\mu ) = \frac{1}{2}\sum\limits_{j = 1}^{\mathop \mu \nolimits_1 } {\mathop \mu \nolimits_j^t (\mathop \mu \nolimits_j^t  - 1)}  = \sum\limits_{s \in \mu } {\mathop {l'}\nolimits_\mu  (s)}  = \sum\limits_{s \in \mu } {\mathop l\nolimits_\mu  (s)} }
\end{eqnarray}
\begin{eqnarray}
\label{A2}
{n(\mathop \mu \nolimits^t ) = \frac{1}{2}\sum\limits_{i = 1}^{d(\mu )} {\mathop \mu \nolimits_i (\mathop \mu \nolimits_i  - 1)}  = \sum\limits_{s \in \mu } {\mathop {a'}\nolimits_\mu  (s)}  = \sum\limits_{s \in \mu } {\mathop a\nolimits_\mu  (s)} }
\end{eqnarray}
\begin{eqnarray}
\label{A3}
{\mathop \kappa \nolimits_\mu   = 2(n(\mathop \mu \nolimits^t  ) - n(\mu)) = \mathop {\left\| \mu \right\|}\nolimits^2  - \mathop {\left\| {\mathop \mu \nolimits^t }  \right\|}\nolimits^2 }
\end{eqnarray}
\begin{eqnarray}
\label{A4}
{\sum\limits_{s \in \mu } {\mathop h\nolimits_\mu  (s)}  = n(\mu ) + n(\mathop \mu \nolimits^t ) + \left| \mu  \right|}
\end{eqnarray}

%%%%%%%%%%%%%%%%%%%%%%%%%%%%%%%%%%%%%%%%%%%%%%%%%%%%%%%%%%%%%%%%%%%%%%%%%%%%%
{\bf Schur functions}

The Schur functions for $N$ variables $(\mathop x\nolimits_1 , \cdots ,\mathop x\nolimits_N )$ are defined by the determinant formula
\begin{eqnarray}
\mathop s\nolimits_\mu  (\mathop x\nolimits_1 , \cdots ,\mathop x\nolimits_N ) = \frac{{\mathop {\det }\nolimits_{i,j = 1, \cdots N} \left( {\mathop {\mathop x\nolimits_i }\nolimits^{\mathop \mu \nolimits_j  + N - j} } \right)}}{{\mathop {\det }\nolimits_{i,j = 1, \cdots N} \left( {\mathop {\mathop x\nolimits_i }\nolimits^{N - j} } \right)}}
\end{eqnarray}
From the definition, the Schur functions are symmetric under the permutation of the variables. Moreover it is known that they form an orthogonal basis of the symmetric polynomials.
We can also define the skew Schur functions by
\begin{eqnarray}
\mathop s\nolimits_{{\mu  \mathord{\left/
 {\vphantom {\mu  \nu }} \right.
 \kern-\nulldelimiterspace} \nu }} (x) = \sum\limits_\rho  {\mathop c\nolimits_{\nu \rho }^\mu  \mathop s\nolimits_\rho  (x)} 
\end{eqnarray}
Here we introduce the Richardson-Littlewood coefficients $\mathop c\nolimits_{\mu \nu }^\rho$
\begin{eqnarray}
\mathop s\nolimits_\mu  (x)\mathop s\nolimits_\nu  (x) = \sum\limits_\rho  {\mathop c\nolimits_{\mu \nu }^\rho  \mathop s\nolimits_\rho  (x)} 
\end{eqnarray}
We have the product expression for the Schur function of the variables $\{ \mathop q\nolimits^\rho  \}  = \mathop {\{ \mathop q\nolimits^{ - i + \frac{1}{2}} \} }\nolimits_{i = 1,2, \cdots } $ \cite{macdonald}
\begin{eqnarray}
\mathop s\nolimits_\mu  (\mathop q\nolimits^{ - \rho } ) \equiv \mathop s\nolimits_\mu  (\mathop q\nolimits^{\frac{1}{2}} ,\mathop q\nolimits^{\frac{3}{2}} , \cdots ) = \mathop q\nolimits^{\frac{{\left\| {\mu ^t } \right\|^2 }}{2}} \mathop {\tilde Z}\nolimits_{\mu }  (q)
\end{eqnarray}
where
\begin{eqnarray}
\mathop {\tilde Z}\nolimits_\mu  (q) = \prod\limits_{s \in \mu } {\left( {1 - \mathop q\nolimits^{\mathop h\nolimits_\mu  (s)} } \right)^{ - 1} } 
\end{eqnarray}
Using this formula, we obtain
\begin{eqnarray}
\mathop s\nolimits_\mu  (\mathop q\nolimits^\rho  ) = \mathop q\nolimits^{\frac{{\mathop \kappa \nolimits_\mu  }}{2}} \mathop s\nolimits_{\mathop \mu \nolimits^t } (\mathop q\nolimits^\rho  ) = \mathop {( - 1)}\nolimits^{\left| \mu  \right|} \mathop s\nolimits_{\mathop \mu \nolimits^t } (\mathop q\nolimits^{ - \rho } )\end{eqnarray}
Let us introduce the 2-parameter extension of $\mathop {\tilde Z}\nolimits_\mu  (q)$ by
\begin{eqnarray}
\mathop P\nolimits_\mu  (\mathop t\nolimits^{ - \rho } ;q,t) = \mathop t\nolimits^{\frac{{\left\| {\mu ^t } \right\|^2 }}{2}} \mathop {\tilde Z}\nolimits_{\mu ^t } (t,q)
\end{eqnarray}
\begin{eqnarray}
\mathop {\tilde Z}\nolimits_\mu  (t,q) = \prod\limits_{s \in \mu } {\left( {1 - \mathop t\nolimits^{\mathop a\nolimits_\mu  (s) + 1} \mathop q\nolimits^{\mathop l\nolimits_\mu  (s)} } \right)^{ - 1} } 
\end{eqnarray}
It appears in the refinement of topological vertex:
\begin{eqnarray}
\mathop C\nolimits_{\phi \phi \mu } (q) = \mathop q\nolimits^{\frac{{\left\| \mu  \right\|^2 }}{2}} \mathop {\tilde Z}\nolimits_\mu  (q) \to \mathop C\nolimits_{\phi \phi \mu } (t,q) = \mathop q\nolimits^{\frac{{\left\| \mu  \right\|^2 }}{2}} \mathop {\tilde Z}\nolimits_\mu  (t,q)
\end{eqnarray}
In summing the Schur functions, we use the following identities
\begin{eqnarray}
\label{sum1}
\sum\limits_\mu  {\mathop s\nolimits_\mu  (x)\mathop s\nolimits_\mu  (y)}  = \prod\limits_{i,j} {(1 - \mathop x\nolimits_i \mathop y\nolimits_j )^{ - 1} } 
\end{eqnarray}
\begin{eqnarray}
\sum\limits_\mu  {\mathop s\nolimits_{\mu ^t } (x)\mathop s\nolimits_\mu  (y)}  = \prod\limits_{i,j} {(1 + \mathop x\nolimits_i \mathop y\nolimits_j )} 
\end{eqnarray}
\begin{eqnarray}\sum\limits_\mu  {\mathop s\nolimits_{\mu /\rho } (x)\mathop s\nolimits_{\mu /\sigma } (y)}  = \prod\limits_{i,j} {\mathop {(1 - \mathop x\nolimits_i \mathop y\nolimits_j )}\nolimits^{ - 1} } \sum\limits_\nu  {\mathop s\nolimits_{\rho /\nu } (y)\mathop s\nolimits_{\sigma /\nu } (x)} 
\end{eqnarray}
\begin{eqnarray}
\sum\limits_\mu  {\mathop s\nolimits_{\mu ^t /\rho } (x)\mathop s\nolimits_{\mu /\sigma } (y)}  = \prod\limits_{i,j} {(1 + \mathop x\nolimits_i \mathop y\nolimits_j )} \sum\limits_\nu  {\mathop s\nolimits_{\rho /\nu ^t } (y)\mathop s\nolimits_{\sigma ^t /\nu ^t } (x)} 
\end{eqnarray}
\begin{eqnarray}
\mathop s\nolimits_\mu  (Qx) = \mathop Q\nolimits^{\left| \mu  \right|} \mathop s\nolimits_\mu  (x)
\end{eqnarray}
\begin{eqnarray}
\mathop s\nolimits_{\mu /\nu } (Qx) = \mathop Q\nolimits^{\left| \mu  \right| - \left| \nu  \right|} \mathop s\nolimits_{\mu /\nu } (x)
\end{eqnarray}
They are important identities which we use throughout the paper.

%%%%%%%%%%%%%%%%%%%%%%%%%%%%%%%%%%%%%%%%%%%%%%%%%%%%%%%%
\section{Proof of Formula }
\label{exam}
%%%%%%%%%%%%%%%%%%%%%%%%%%%%%%%%%%%%%%%%%%%%%%%%%%%%%%%%

In this appendix, we prove the following identity for section 3.
\begin{align}
\label{b}
&\prod\limits_{a = 1}^N {\mathop {\mathop Q\nolimits_{M,a} }\nolimits^{\left| {\mathop \mu \nolimits_a } \right|} } \prod\limits_{a = 2}^{N } {\mathop {\mathop Q\nolimits_{F,a} }\nolimits^{\left| {\mathop \nu \nolimits_a } \right|} } \sum\limits_{
\scriptsize \{\mu_i\},~\{\nu_i\}\atop
\scriptsize \{\rho_i\},~\{\sigma_i\}
} 
{\prod\limits_{a = 1}^N {\mathop s\nolimits_{\mathop \nu \nolimits_a /\mathop \rho \nolimits_a } (\mathop w\nolimits^{(a)} )\mathop s\nolimits_{\mathop \mu \nolimits_a /\mathop \rho \nolimits_a } (\mathop x\nolimits^{(a)} )\mathop s\nolimits_{\mathop {\mathop \mu \nolimits_a }\nolimits^t /\mathop \sigma \nolimits_{a + 1} } (\mathop y\nolimits^{(a)} )\mathop s\nolimits_{\mathop {\mathop \nu \nolimits_{a + 1} }\nolimits^t /\mathop \sigma \nolimits_{a + 1} } (\mathop z\nolimits^{(a + 1)} )} } 
 \cr
 &= \prod\limits_{1 \le a \le b \le N} {\left[ {\mathop x\nolimits^{(a)} ,\mathop Q\nolimits_{\mathop \alpha \nolimits_a \mathop \beta \nolimits_b } \mathop y\nolimits^{(b)} } \right]} \prod\limits_{1 \le a < b \le N} {\left[ {\mathop z\nolimits^{(a+1)} ,\mathop Q\nolimits_{\mathop \beta \nolimits_a \mathop \alpha \nolimits_b } \mathop w\nolimits^{(b)} } \right]} \left\{ {\mathop x\nolimits^{(a)} ,\mathop Q\nolimits_{\mathop \alpha \nolimits_a \mathop \alpha \nolimits_b } \mathop w\nolimits^{(b)} } \right\}\left\{ {\mathop z\nolimits^{(a+1)} ,\mathop Q\nolimits_{\mathop \beta \nolimits_a \mathop \beta \nolimits_b } \mathop y\nolimits^{(b)} } \right\}
\end{align}
where we take the sum over the Young diagrams $\mathop \mu \nolimits_1  \cdots \mathop \mu \nolimits_N $, $\mathop \nu \nolimits_2  \cdots \mathop \nu \nolimits_N 
$, $\mathop \rho \nolimits_2  \cdots \mathop \rho \nolimits_N $, and $\mathop \sigma \nolimits_2  \cdots \mathop \sigma \nolimits_N $. Notice that we denote
$\mathop \rho \nolimits_1  = \mathop \sigma \nolimits_{N + 1}  = \phi $. in the formula.

Let us show the identity. The first line of this equation becomes
\begin{align}
&\sum\limits_{
   {\mathop \rho \nolimits_2  \cdots \mathop \rho \nolimits_N }  \atop
   {\mathop \sigma \nolimits_2  \cdots \mathop \sigma \nolimits_N }  
} {\prod\limits_{a = 2}^N {\mathop {\mathop Q\nolimits_{M,a} }\nolimits^{\left| {\mathop \rho \nolimits_a } \right|} \mathop {\mathop Q\nolimits_{F,a} }\nolimits^{\left| {\mathop \sigma \nolimits_a } \right|} } }  \cr
&\times \sum\limits_{
   {\mathop \mu \nolimits_1  \cdots \mathop \mu \nolimits_N }  \atop
   {\mathop \nu \nolimits_2  \cdots \mathop \nu \nolimits_N }  
} {\prod\limits_{a = 1}^N {\mathop s\nolimits_{\mathop \mu \nolimits_a /\mathop \rho \nolimits_a }  (\mathop Q\nolimits_{M,a} \mathop x\nolimits^{(a)} )\mathop s\nolimits_{\mathop {\mathop \mu \nolimits_a }\nolimits^t /\mathop \sigma \nolimits_{a + 1} } (\mathop y\nolimits^{(a)} )\mathop s\nolimits_{\mathop {\mathop \nu \nolimits_{a + 1} }\nolimits^t /\mathop \sigma \nolimits_{a + 1} } (\mathop Q\nolimits_{F,a + 1} \mathop z\nolimits^{(a + 1)} )\mathop s\nolimits_{\mathop \nu \nolimits_{a + 1} /\mathop \rho \nolimits_{a + 1} } (\mathop w\nolimits^{(a + 1)} )} } \nonumber
\end{align}
\begin{align}
&= \prod\limits_{a = 1}^N {\left[ {\mathop x\nolimits^{(a)} ,\mathop Q\nolimits_{M,a} \mathop y\nolimits^{(a)} } \right]} \prod\limits_{a = 2}^N {\left[ {\mathop z\nolimits^{(a + 1)} ,\mathop Q\nolimits_{F,a + 1} \mathop w\nolimits^{(a + 1)} } \right]} 
\sum\limits_{
   {\mathop \alpha \nolimits_2  \cdots \mathop \alpha \nolimits_{N - 1} }  \atop   {\mathop \beta \nolimits_2  \cdots \mathop \beta \nolimits_N }  
} {\sum\limits_{
   {\mathop \rho \nolimits_2  \cdots \mathop \rho \nolimits_N }  \atop
   {\mathop \sigma \nolimits_2  \cdots \mathop \sigma \nolimits_N }  
} {\prod\limits_{a = 2}^N {\mathop {\mathop Q\nolimits_{M,a} }\nolimits^{\left| {\mathop \alpha \nolimits_a } \right|} \mathop {\mathop Q\nolimits_{F,a} }\nolimits^{\left| {\mathop \beta \nolimits_a } \right|} } } }  
  \cr
&\times \prod\limits_{a = 1}^N {\mathop s\nolimits_{\mathop {\mathop \sigma \nolimits_{a + 1} }\nolimits^t /\mathop \alpha \nolimits_a } (\mathop Q\nolimits_{M,a} \mathop x\nolimits^{(a)} )\mathop s\nolimits_{\mathop {\mathop \rho \nolimits_a }\nolimits^t /\mathop \alpha \nolimits_a } (\mathop Q\nolimits_{M,a} \mathop y\nolimits^{(a)} )\mathop s\nolimits_{\mathop {\mathop \rho \nolimits_{a + 1} }\nolimits^t /\mathop {\mathop \beta \nolimits_{a + 1} }\nolimits^t } (\mathop Q\nolimits_{F,a + 1} \mathop z\nolimits^{(a + 1)} )\mathop s\nolimits_{\mathop {\mathop \sigma \nolimits_{a + 1} }\nolimits^t /\mathop \beta \nolimits_{a + 1} } (\mathop Q\nolimits_{F,a + 1} \mathop w\nolimits^{(a + 1)} )} \nonumber
\end{align}

\begin{align}
&= \prod\limits_{a = 1}^N {\left[ {\mathop x\nolimits^{(a)} ,\mathop Q\nolimits_{M,a} \mathop y\nolimits^{(a)} } \right]} \prod\limits_{a = 2}^{N - 1} {\left[ {\mathop z\nolimits^{(a)} ,\mathop Q\nolimits_{F,a} \mathop w\nolimits^{(a)} } \right]} \prod\limits_{a = 1}^N {\left\{ {\mathop x\nolimits^{(a)} ,\mathop Q\nolimits_{M,a} \mathop Q\nolimits_{F,a + 1} \mathop w\nolimits^{(a + 1)} } \right\}\left\{ {\mathop z\nolimits^{(a + 1)} ,\mathop Q\nolimits_{F,a + 1} \mathop Q\nolimits_{M,a + 1} \mathop y\nolimits^{(a + 1)} } \right\}} \cr
 &\times  \sum\limits_{
   {\mathop \beta \nolimits_1  \cdots \mathop \beta \nolimits_{N - 1} }  \atop
   {\mathop \alpha \nolimits_2  \cdots \mathop \alpha \nolimits_{N - 1} }  
} {\sum\limits_{
   {\mathop \gamma \nolimits_2  \cdots \mathop \gamma \nolimits_{N - 1} }  \atop   {\mathop \delta \nolimits_2  \cdots \mathop \delta \nolimits_{N - 1} }  
} {\prod\limits_{a = 1}^{N - 1} {\mathop {\mathop Q\nolimits_{F,a + 1} }\nolimits^{\left| {\mathop \beta \nolimits_a } \right|} } \prod\limits_{a = 2}^{N - 1} {\mathop {\mathop Q\nolimits_{M,a} }\nolimits^{\left| {\mathop \alpha \nolimits_a } \right|} } } }  \cr
&\times  \prod\limits_{a = 1}^{N - 1} {\mathop s\nolimits_{\mathop \alpha \nolimits_a /\mathop \gamma \nolimits_a } (\mathop Q\nolimits_{F,a + 1} \mathop w\nolimits^{(a + 1)} )\mathop s\nolimits_{\mathop \beta \nolimits_a /\mathop \gamma \nolimits_a } (\mathop Q\nolimits_{M,a} \mathop x\nolimits^{(a)} )\mathop s\nolimits_{\mathop {\mathop \beta \nolimits_a }\nolimits^t /\mathop \delta \nolimits_{a + 1} } (\mathop Q\nolimits_{M,a + 1} \mathop y\nolimits^{(a + 1)} )\mathop s\nolimits_{\mathop {\mathop \alpha \nolimits_{a + 1} }\nolimits^t /\mathop \delta \nolimits_{a + 1} } (\mathop Q\nolimits_{F,a + 1} \mathop z\nolimits^{(a + 1)} )} \notag
\end{align}
Using this result repeatingly, we obtain the second line of the formula (\ref{b}).

\end{document}